\documentclass[11pt]{article}
\usepackage{booktabs,multirow}
\usepackage[linesnumbered,ruled]{algorithm2e}
\usepackage{graphicx}
\usepackage{dsfont}
\usepackage{grffile}
\usepackage{longtable,multicol}
\usepackage{wrapfig}
\usepackage{rotating}
\usepackage[normalem]{ulem}
\usepackage{amsfonts,amsmath,mathtools}
\usepackage{textcomp}
\usepackage{amssymb}
\usepackage{capt-of}
\usepackage{hyperref}
\usepackage{bbm}
\usepackage{setspace}
\usepackage{comment}
\usepackage[bindingoffset=1.25cm,left=1.5cm,right=2cm,top=2cm,bottom=2cm, footskip=1cm]{geometry}
\usepackage[T1]{fontenc}
\usepackage[varqu,varl,var0]{inconsolata}
\usepackage[scale=.95,type1]{cabin}
\usepackage{natbib}
\usepackage{hyperref}
\hypersetup{colorlinks=true,linkcolor=blue,citecolor=blue}

\usepackage{subcaption}
\usepackage[edges]{forest}
\forestset{default preamble={for tree={draw}}}

\usepackage{todonotes}
\newcounter{todocounter}

\newtheorem{assumption}{Assumption}[section]

\newcommand{\indep}{\perp \!\!\! \perp}
\newcommand{\B}{\mathcal{B}}
\author{Rafael Alcantara \thanks{The University of Texas at Austin} \and Meijia Wang \thanks{Google} \and P. Richard Hahn \thanks{Arizona State University} \and Hedibert Lopes \thanks{Insper}}
\title{Modified BART for Learning Heterogeneous Effects\\ in Regression Discontinuity Designs}
\begin{document}
\maketitle
\begin{abstract}
	This paper introduces BART-RDD, a sum-of-trees regression model built around a novel regression tree prior, which incorporates the special covariate structure of regression discontinuity designs. Specifically, the tree splitting process is constrained to ensure overlap within a narrow band surrounding the running variable cutoff value, where the treatment effect is identified. It is shown that unmodified BART-based models estimate RDD treatment effects poorly, while our modified model accurately recovers treatment effects at the cutoff. Specifically, BART-RDD is perhaps the first RDD method that effectively learns conditional average treatment effects. The new method is investigated in thorough simulation studies as well as an empirical application looking at the effect of academic probation on student performance in subsequent terms \citep{lindo2010ability}.
\end{abstract}
\textbf{Keywords:} Bayesian causal forest, tree ensembles, regression discontinuity design
\newcommand{\res}{\mathbf{r}}
\newcommand{\w}{\mathbf{w}}
\newcommand{\m}{\mathbf{m}}
\newcommand{\x}{\mathbf{x}}
\newcommand{\C}{\mathbb{C}}
\newcommand{\E}{\mathds{E}}
\newcommand{\N}{\mathrm{N}}
\SetKwComment{Comment}{/* }{ */}
\section{Introduction}
\label{sec:introduction}
Regression discontinuity designs (RDD), originally proposed by \citet{thistlethwaite1960regression}, are widely used in economics and other social sciences to estimate treatment effects from observational data. Such designs arise when treatment assignment is based on whether a particular covariate --- referred to as the running variable --- lies above or below a known value, referred to as the cutoff value. Thus in an RDD, deterministic treatment assignment implies that conditional confoundedness is trivially satisfied, given the running variable. However, controlling for the running variable introduces a complete lack of overlap. Thus, identification of treatment effects from RDDs relies on assumptions that permit coping with this lack of overlap. First, individuals are unable to manipulate their realization of the forcing variable. For example, if the forcing variable is a test score and students can take the test only once, then students who score slightly above or below the cutoff are likely very similar except for their position relative to the cutoff. On the other hand, if students could retake the test arbitrarily often, then a student who scored above the cutoff on their first try and a student who did so only on their tenth try are most likely not similar, but both would be eligible for treatment. Second, the conditional expectation of the response variable, given the forcing variable, must be smooth at the cutoff point. Without this assumption, it would not possible to disambiguate between a treatment effect and non-causal aspects of the data-generating process \citep{lee2010regression,imbens2008regression}.


Previous work has shown that treatment effects can be estimated from RDDs as the magnitude of a discontinuity in the conditional mean response function at the cutoff \citep{hahn2001identification}. This paper investigates the use of Bayesian additive regression tree models \citep{chipman2010bart, hahn2020bayesian} for the purpose of fitting RDD data with additional covariates for estimating conditional average treatments effects (CATE) at the cutoff. Broadly, our work expands on both frequentist and Bayesian methods for performing RDD analyses that incorporate covariates in addition to the running variable. Relative to earlier works, the method proposed here accommodates a richer set of data generating processes and admits more convenient sensitivity analysis and methods for visualizing heterogeneity. Most importantly, our estimator is one of the few RDD estimators which allow for exploring heterogeneity in a data-driven manner, instead of relying on separate ATE estimation for predetermined subgroups. 

\subsection{Previous work}
The inclusion of covariates in RDD models has been studied
by \cite{calonico2019regression}, who extend the local linear
regression to include covariates linearly and discuss the
implications of this in terms of bias and variance, and
\cite{frolich2019including} who propose a nonparametric kernel
regression method which increases precision and may reduce
bias and restore identification in data with discontinuities
in the covariate set at the threshold, provided that all
relevant discontinuous covariates are
included. Additionally, \cite{arai2021regression} and
\cite{kreiss2023inference} study RDD regressions with
high-dimensional covariate sets. The latter two essentially
consist of a pre-selection step where one fits a variable
selection model (typically a Lasso) to either the full
sample or a subsample closer to the cutoff and then fits the
local polynomial estimator of \citet{calonico2019regression}
to the reduced feature set. These methods inherit the local
polynomial's linearity assumption which can lead to high
variance estimators in the presence of strong heterogeneity
if the running variable and other features interact in more
complex ways. The estimator proposed by
\citet{frolich2019including} is more flexible in that regard
because it allows for feature-specific kernels, extending
the traditional local RDD regression. However, this can
render the method computationally infeasible as the
dimension of the feature set increases.

The previous works discuss inclusion of covariates from a
perspective of obtaining precision gains and barely discuss
effect moderation. Regarding effect heterogeneity,
\cite{frandsen2012quantile} and \cite{shen2016distributional} discuss it
from the perspective of distributional effects, while
\cite{cattaneo2016interpreting} and \cite{bertanha2020regression} discuss
heterogeneity arising from settings with multiple
cutoffs. These works do not focus specifically on
heterogeneity in the form of moderation by additional
variables. \cite{becker2013absorptive} extend the traditional
local regression to include interaction terms between the
running variable and additional
features. \cite{hsu2019testing} develop hypothesis tests for
detecting effect moderation in the local regression setup by
means of comparison between the ATE parameter for the whole
sample and pre-specified subgroups of interest. These
methods still depend on reasonable previous knowledge about
potential sources of heterogeneity.

Prominent examples of Bayesian estimators for RDDs, include
\cite{chib2023nonparametric}, who estimate the response curves
with global splines where observations are weighted by their
distance to the cutoff; \cite{karabatsos2015bayesian}, who
propose approximating the conditional expectations by an
infinite mixture of normals; and
\cite{branson2019nonparametric}, who propose a Gaussian
process prior for the expectations, in which observations
are also weighted by their distance to the cutoff. All of
these methods consist of global approximations of the
outcome curves, while in some cases emphasizing units near
the cutoff to obtain better predictions in that region. As
will be discussed later, our method can be seen as an
intermediate approach between such global approximations and
the local linear regression ubiquitous in the frequentist
literature, since we use the entire data to estimate the
outcomes but take advantage of the local nature of BART for
estimation near the cutoff.

\cite{reguly2021heterogeneous} proposes what is perhaps the
closest in spirit to our method. The author proposes a
modification to the basic Classification and Regression Tree
(CART) algorithm in which the tree is split using all
features available except for the running variable. Then,
within each leaf the algorithm performs a separate
regression for treated and untreated units, and the
leaf-specific ATE parameter is obtained as the difference
between the intercepts of the two regressions. The model can
be seen as a non-linear polynomial regression where the
parameters depend on the covariates via the CART fit. This
approach presents two important differences compared to
ours. First, it is a single tree method, whereas we propose
a forest model. Second, the flexibility of the tree is only
used for the additional covariates, but the leaf regressions
are still polynomials of the running variable. These
features mean that, although more flexible than global
regressions, the method should still suffer in situations
where the response surface is smoother on the covariates or
where the way the running variable interacts with the others
is more complex than a polynomial model could accurately
capture. Still, this is, to the best of our knowledge, the
only RDD estimator beside our own which does not require
pre-specification of subgroups for CATE analysis.

\section{Background}
This paper brings together ideas from many different areas, each with their own terminology and notation. In this section we review the basics of regression discontinuity designs, BART, and Bayesian causal forests.
\subsection{Regression Discontinuity Designs}
\label{sec:rdd}
Following \citet{imbens2008regression}, we frame the RD setting in a potential outcomes model, which can be briefly described as follows. Let \(Z\) denote a binary treatment
variable and \(Y_i^z\) denote the potential outcome of unit \(i\) under treatment state \(Z_i=z\). The treatment effect for
unit \(i\) is defined as:
\begin{equation}
	\tau_i \coloneqq Y_i^1-Y_i^0.
\end{equation}
Let \(X\) denote the running variable and \(W\) denote a set of additional covariates. Commonly, interest lies on the average and conditional average treatment effect (ATE/CATE):
\begin{equation}
	\begin{split}
		\E[\tau_i] &= \E[Y_i^1-Y_i^0] \\
		\E[\tau_i \mid X,W] &= \E[Y_i^1-Y_i^0 \mid X,W].
	\end{split}
\end{equation}
These quantities are of course unobservable since each unit
is only observed at a single given treatment state. However,
under the following assumptions we can link \(\tau\) to the
observed outcome and covariates \((Y,Z,X,W)\).

\begin{assumption}[SUTVA]
	\label{ass:sutva}
	This assumption has two components: consistency and
	no-interference, which are represented, respectively, as:
	\begin{equation}
		\begin{split}
			&Y = Y^0 + Z(Y^1 - Y^0) \label{eq:y}\\
			&Y^1_i,Y^0_i \indep Z_j,
		\end{split}
	\end{equation}
	for all $i,j \in \{1,\dots,n\}$, where $i \neq j$.
\end{assumption}

\begin{assumption}[Mean conditional unconfoundedness]
	\label{ass:unconf}
	$Y^1,Y^0$ are mean independent of $Z$ conditional on $X,W$:
	\begin{equation}
		\begin{split}
			\E[Y^1 \mid Z,X,W] &= E[Y^1 \mid X,W]\\
			\E[Y^0 \mid Z,X,W] &= E[Y^0 \mid X,W].
		\end{split}
	\end{equation}
\end{assumption}

Under assumptions \eqref{ass:sutva} and \eqref{ass:unconf}, the CATE is identified as:
\begin{equation}
	\E[Y^1-Y^0 \mid X,W] = \E[Y \mid Z=1,X,W] - \E[Y \mid Z=0,X,W].
\end{equation}

While the previous assumptions lead to identification of the CATE, one more assumption is necessary for estimation:

\begin{assumption}[Conditional overlap]
	\label{ass:overlap}
	Both treatment states have a positive probability
	conditional on $X,W$:
	\begin{equation}
		0 < P(Z=z \mid X,W) < 1.
	\end{equation}
\end{assumption}

In words, assumption \eqref{ass:overlap} allows one to compare treated and untreated units in any region of the support of $(X,W)$, leading to the construction of valid causal contrasts.

The distinctive feature of the RDD is that \(Z\) is a
deterministic function of \(X\):
\begin{equation*}
	Z_i= \left\{
	\begin{array}{ll}
		0, \quad \text{if} \quad X_i &< c \\
		1, \quad \text{if} \quad X_i &\geq c
	\end{array}
	\right.
\end{equation*}
for a known cutoff value \(c\)\footnote{Our focus lies on the so-called ``sharp'' RDD, in
	which case there is perfect compliance --- as opposed to the
	``fuzzy'' RDD, in which case compliance is imperfect --- so
	the perfect compliance assumption is implicit throughout the
	text.}.

The deterministic assignment mechanism implies that controlling for \(X\) is sufficient to ensure unconfoundedness. However, this control induces a complete lack of overlap. Therefore, treatment effect estimation in the RDD requires additional assumptions to circumvent this issue. In order to discuss these assumptions, we write the expectation of $Y$ given $(X,W,Z)$ as:

\begin{equation}
	\begin{split}
		\E[Y \mid X,W,Z] &= \mu(X,W) + \tau(X,W) Z\\
		\mu(X,W) &= \E[Y \mid X, W, Z=0]\\
		\tau(X,W) &= \E[Y \mid X, W, Z=1] - \E[Y \mid X, W, Z=0].
	\end{split}
\end{equation}

Because of the lack of overlap, one can only learn $\mu(X,W)$ in the region $X < c$ and $\mu(X,W) + \tau(X,W)$ in the region $X \geq c$, so that inferences concerning $\tau(X,W)$ at arbitrary $x$ cannot be obtained without further assumptions. We now discuss the kinds of assumptions necessary for estimation in the RDD.

For some $\epsilon > 0$, let $x_{\varepsilon}^- = \{x \in (c-\epsilon,c)\}$, $x_{\varepsilon}^+ = \{x \in [c,c+\epsilon)\}$, and $x_{\epsilon} = x_{\epsilon}^- \cup x_{\epsilon}^+$. Suppose that, for $X \in x_{\epsilon}$, the treatment effect function is independent from the treatment variable conditional on $X$. Then:

\begin{equation}
	\begin{split}
		&\E[\E[Y \mid X,W,Z=1] \mid X \in x_{\epsilon}] - \E[\E[Y \mid X,W,Z=0] \mid X \in x_{\epsilon}]\\
		= &\E[\tau(X,W) \mid X \in x_{\epsilon}^+] + (\E[\mu(X,W) \mid X \in x_{\epsilon}^+] - \E[\mu(X,W) \mid X \in x_{\epsilon}^-]).
	\end{split} 
\end{equation}

Suppose that:
\begin{equation} \label{eq:rdd.ate}
	\begin{split}
		\E[\mu(X,W) \mid X \in x_{\epsilon}^+] &= \E[\mu(X,W) \mid X \in x_{\epsilon}^-] = \E[\mu(X,W) \mid X \in x_{\epsilon}]\\
		\E[\tau(X,W) \mid X \in x_{\epsilon}^+] &= \E[\tau(X,W) \mid X \in x_{\epsilon}^-] = \E[\tau(X,W) \mid X \in x_{\epsilon}].
	\end{split}
\end{equation}

Then, the ATE\footnote{As is commonly done in the RDD literature, we refer to the CATE conditional only on $X$ as the ATE and use CATE only when conditioning on $W$ as well} is identified inside this region:

\begin{equation}
	\begin{split}
		&\E[\E[Y \mid X,W,Z=1] \mid X \in x_{\epsilon}] - \E[\E[Y \mid X,W,Z=0] \mid X \in x_{\epsilon}]\\
		= &\E[\tau(X,W) \mid X \in x_{\epsilon}].
	\end{split} 
\end{equation}

This is the basis of the continuity-based identification approach introduced by \cite{hahn2001identification}. Under that setting, if these conditions can be assumed to hold at least as $\epsilon \to 0$ --- \textit{i.e.} if the expectation of the $\mu$ and $\tau$ functions are continuous at $X=c$ --- the ATE at this point is identified as $\E[\tau(X=c,W)]$.

If interest lies in identification of the CATE in some region of the feature set $W=w$, we need similar assumptions about the expectations conditional on $W$. Suppose that, for all $x_{-} \in x_{\epsilon}^-$ and $x_{+} \in x_{\epsilon}^+$:

\begin{equation} \label{eq:rdd.cate}
	\begin{split}
		\mu(X=x_{-},W=w) &= \mu(X=x_{+},W=w)\\
		\tau(X=x_{-},W=w) &= \tau(X=x_{+},W=w).
	\end{split}
\end{equation}

Then, the CATE at $W=w$ is identified in the region $x_{\epsilon}$ by $\tau(X,W=w)$. As before, if these equalities hold as $\epsilon \to 0$, \textit{i.e.} if $\mu$ and $\tau$ are continuous at $X=c$, the CATE for $W=w$ is identified at that point. Because we propose an estimator for the CATE, \eqref{eq:rdd.cate} is assumed to hold for the remainder of the text. However, it is worth emphasizing that only \eqref{eq:rdd.ate} is required for identification of the ATE, so that, even if $\tau(X,W)$ does not identify any CATE, estimates of this function can still be used to produce ATE estimates if the relevant assumptions hold.

To introduce some of the challenges faced by tree models in the RDD context, consider the treatment effect estimate in a single tree model for a partition in the tree fit that contains $X=c$, denoted by $\B$. For points inside that partition, define $X^\B_{+} = x \in [c,\overline{x}]$, $X^\B_{-} = x \in [\underline{x},c]$, where $\underline{x}$ and $\overline{x}$ are the smallest and largest values of $X$ inside the partition, respectively, and $X^\B = X^\B_{+} \cup X^\B_{-}$. Then:

\begin{equation}
	\begin{split}
		&E[Y \mid X \in X^\B,W,Z=1] - E[Y \mid X \in X^\B,W,Z=0]\\
		= &\mu(X \in X^\B_{+},W)-\mu(X \in X^\B_{-},W) + \tau(X \in X^\B_{+},W).
	\end{split}
\end{equation}

This means that the bias for the cutoff treatment effect estimate inside this partition is given by:

\begin{equation}
	\tau_{\text{bias}} = \tau(X=c,W) - \tau(X \in X^\B_{+},W) + \mu(X \in X^\B_{+},W)-\mu(X \in X^\B_{-},W). \label{eq:bias}
\end{equation}

Equation \eqref{eq:bias} shows how the bias in a tree model is determined by the composition of the leaf nodes. In particular, it implies that the bias goes to zero as $\underline{x} \to c$ and $\overline{x} \to c$. In words, although nodes that are too tight around $X=c$ could lead to an increase in variance due to the decreasing number of available points in the leaves, nodes that contain too wide regions around the cutoff could lead to extremely biased estimates if $\mu$ and $\tau$ feature a wide range of values in that partition. Therefore, when considering a split in a tree, minimal variation of the prognostic and treatment effect functions around the cutoff inside the generated leaves should be a key component of the tree growth process. This is the essence of the BART-RDD model, which will be discussed in more detail in section \ref{sec:xbcf.rdd}.

\subsection{Bayesian Additive Regression Trees}
\label{sec:bart}
The Bayesian Additive Regression Trees model
\citep{chipman2010bart}, or BART, represents an unknown mean
function as a sum of regression trees, where each regression
tree is assumed to be drawn from the tree prior described in
\cite{chipman1998bayesian}. Letting \(f(x) = \E(Y \mid X = x)\)
denote a smooth function of a covariate vector \(X\), the BART
model is traditionally written
\begin{equation} \label{eq:bart.model}
	\begin{split}
		Y_i &= f(x_i) + \varepsilon_i,\\
		&= \sum_{j=1}^k g_j(x_i; T_j, \m_j) + \varepsilon_i,\\
	\end{split}
\end{equation}
where \(\varepsilon_i \sim \N(0,\sigma^2)\) is a normally
distributed additive error term. Here, each \(g_j(x ; T_j,
\m_j)\) denotes a piecewise function of \(x\) defined by a set
of splitting rules \(T_j\) that partitions the domain of \(x\)
into disjoint regions, and a vector, \(\m_j\), which records
the values \(g(\cdot)\) takes on each of those
regions. Therefore, the parameters of a standard BART
regression model are \((T_1, \m_1), \dots, (T_k, \m_k)\) and
\(\sigma\). \cite{chipman2010bart} consider priors such that:
the tree components \((T_j, \m_j)\) are independent of each
other and of \(\sigma^2\), and the terminal node parameters
\(\mu_{k1}, \dots, \mu_{kb}\) of a given tree \(k\) are
independent of each other. Furthermore, \cite{chipman2010bart}
consider the same priors for all trees and leaf node
parameters. The model thus consists of the specification of
three priors: \(p(T)\), \(p(\sigma^2)\) and \(p(\m|T)\).

The tree prior, \(p(T)\), is defined by three
components. First, the probability that a node \(d\) will
split is determined by
\begin{equation}
	\frac{\alpha}{(1+d)^\beta}, \quad \alpha \in (0,1), \beta \in [0,\infty) \label{eq:tree.prior.1}.
\end{equation}
That is, the deeper the node (higher \(d\)), the higher the
chance that it is a terminal node. This is essentially a
regularization component of the tree prior to avoid
overfitting.

The other components of the tree prior are the probability
that a given variable will be chosen for the splitting rule
at node \(d\), and the probability that a given observed value
of the chosen variable will be used for the splitting
rule. The splitting variable is chosen uniformly among the
set of covariates and then the splitting value is chosen
uniformly among the discrete set of observed values of that
covariate.

For further details and justification concerning BART prior specification, see \cite{chipman2010bart}.
%
\subsection{Bayesian Causal Forest}
There are two common strategies for estimating heterogeneous
treatment effects. One is to simply focus on estimating the
response surface including a treatment indicator as an
additional covariate, while the other consists of fitting
two different models for treatment and control
groups. Recently, these approaches have been dubbed
``S-learners'' and ``T-learners'' respectively, where S
means ``single'' and T means ``two''
\citep{kunzel2019metalearners}. In the context of applying BART
for causal inference, \cite{hill2011bayesian} follows the
first approach, under the assumption of no unobserved
confounding, which implies that treatment effect estimation
reduces to response surface estimation. For the second
approach, one could simply fit two different BART models for
treated and control units.

As described in \cite{hahn2020bayesian}, neither of these
approaches is ideal in common causal inference settings. The
two-model T-learner approach has the problem that
regularization of the treatment effect is necessarily weaker
than regularization of each individual model, which is the
opposite of what you would expect in many contexts, where
treatment effects are expected to be modest. The
single-model approach of \cite{hill2011bayesian} addresses
this to some extent, but at the expense of transparency: the
implied degree of regularization depends sensitively on the
joint distribution of the control variables and the
treatment variable. Accordingly, \cite{hahn2020bayesian}
proposed the Bayesian Causal Forest (BCF) model, which fits
two BART models simultaneously to a reparametrized response
function:
\begin{equation} \label{eq:bcf.model}
	\begin{split}
		Y_i &= \mu(X_i,\mathrm{w}_i) + \tau(X_i,\mathrm{w}_i) b_{z_i} + \varepsilon_i, \quad \varepsilon_i \sim N(0,\sigma^2),\\
		b_0 &\sim \N(0,1/2), \quad b_1 \sim \N(0,1/2).
	\end{split}
\end{equation}
where \(\mu(\cdot)\) is referred to as a prognostic function
and \(\tau(\cdot)\) a treatment effect function\footnote{This terminology is motivated by the case where \(b_0
	= 0\) and \(b_1 = 1\), in which case \(\mu(x) = \E(Y^0 \mid X =
	x)\) and \(\tau(x) = \E(Y^1 \mid X = x) - \E(Y^0 \mid X = x)\).}. The
model parametrized in this way can be seen as a linear
regression with a covariate-dependent slope and intercept
\citep{hahn2020bayesian}.

Note that \(b_0\) and \(b_1\) are parameters that can be
estimated; practically this is desirable because it avoids
giving the treated potential outcome higher prior predictive
variance\footnote{Treatment coding can imply non-equivalent priors for
	the treatment effects. For example, \(Z \in (0,1)\) implies
	the expected potential outcomes for the treated group depend
	on the treatment effect function, while the expected
	potential outcomes for the control group do not, which is
	not reasonable if we have a comparison between two levels of
	treatment instead of treatment \textit{vs.} no
	treatment. See \cite{hahn2020bayesian} for a more thorough
	discussion on how treatment effect priors are dependent on
	treatment encoding.}. Under this parameterization the treatment
effect can be expressed as:
\begin{equation}
	\E(Y^1 \mid X = x) - \E(Y^0 \mid X = x) = (b_1-b_0) \tau(x).
\end{equation}
\subsection{The XBCF model}
\label{subsec:model}
To sample from the posterior distributions of trees,
\cite{chipman2010bart} propose a backfitting MCMC algorithm
that explores the tree space by proposing at each iteration
a grow or prune step, producing highly correlated tree
samples. This can make convergence of the algorithm slow and may not scale well to large datasets. And, as BCF
depends on BART priors, it will also be affected by these
problems.

As a more efficient alternative, \cite{he2023stochastic}
propose the accelerated Bayesian additive regression trees
(XBART) algorithm for BART-like models. XBART grows new
trees recursively, but stochastically, at each step while
using a similar set of cutpoints and splitting criteria as
BART, which allows for much faster exploration of the
posterior space. \cite{krantsevich2023stochastic} extends the XBART algorithm to the BCF model, with their
accelerated Bayesian causal forest (XBCF) algorithm, an
XBCF algorithm. The XBCF algorithm uses a slightly modified BCF model, allowing the error variance to differ by treatment status:
\begin{equation}
	\begin{split}
		Y_i &= a \mu (x_i) + b_{z_i}\tilde{\tau}(x_i) + \epsilon_i, \quad \epsilon_i \sim N(0, \sigma_{z_i}^2) \\
		a &\sim \N(0, 1), \quad b_0, b_1 \sim \N(0, 1/2),
	\end{split}
\end{equation}
where \(\mu(x)\) and \(\tilde{\tau}(x)\) are two XBART forests
and \(\tau = (b_1-b_0)\tilde{\tau}\).

The BART-RDD method described below is a modified XBCF
algorithm, specialized in critical ways to the RDD setting. The key innovation from \cite{he2023stochastic} is the
so-called ``Grow-From-Root'' stochastic tree-fitting algorithm, reproduced in algorithm \ref{alg:gfr} in a summary form. It will be
discussed later how this algorithm is particularly well-suited to the RDD context.

\begin{algorithm}
	\caption{GrowFromRoot}\label{alg:gfr}
	\KwOut{Modifies T by adding nodes and sampling associated leaf parameters $\mu$.}
	\If{the stopping conditions are met}{Go to step 13, update leaf parameter $\mu_{node}$\;}
	$s^{\emptyset} \gets s(y,\mathbf{X},\mathbf{\Psi},\mathcal{C},\text{all})$\;
	\For{$c_{jk} \in \mathcal{C}$}{$s^{(1)}_{jk} \gets s(y,\mathbf{X},\mathbf{\Psi},\mathcal{C},j,k,\text{left})$\;
		$s^{(2)}_{jk} \gets s(y,\mathbf{X},\mathbf{\Psi},\mathcal{C},j,k,\text{right})$\;
		Calculate $L(c_{jk}) = m \left( s^{(1)}_{jk};\mathbf{\Phi},\mathbf{\Psi} \right) \times m \left( s^{(2)}_{jk};\mathbf{\Phi},\mathbf{\Psi} \right)$\;}
	Calculate $L(\emptyset) = |\mathcal{C}| \left( \frac{(1+d)^{\beta}}{\alpha} - 1 \right) m \left( s^{\emptyset};\mathbf{\Phi},\mathbf{\Psi} \right)$\;
	Sample a cutpoint $c_{jk}$ proportional to integrated likelihoods
	\begin{equation*}
		P(c_{jk}) = \frac{L(c_{jk})}{\sum_{c_{jk} \in \mathcal{C}} L(c_{jk}) + L(\emptyset)},
	\end{equation*}
	or
	\begin{equation*}
		P(\emptyset) = \frac{L(\emptyset)}{\sum_{c_{jk} \in \mathcal{C}} L(c_{jk}) + L(\emptyset)}
	\end{equation*}
	for the null cutpoint\;
	\eIf{the null cutpoint is selected}{$\mu_{node} \gets SampleParameters(\emptyset)$\;
		\textbf{return}}{Create two new nodes, \textbf{left\_node} and \textbf{right\_node}, and grow $T$ by designating them as the current node's (\textbf{node}) children\;
		Partition the data $(y, \mathbf{X})$ into left $(y_{left}, \mathbf{X}_{left})$ and right $(y_{right}, \mathbf{X}_{right})$ parts, according to the selected cutpoint $x_{ij'} \leq x_{jk}^*$ and $x_{ij'} > x_{jk}^*$, respectively, where $x_{jk}^*$ is the value corresponding to the sampled cutpoint $c_{jk}$\;
		$\text{GrowFromRoot}(y_{\text{left}},\mathbf{X}_{\text{left}},\mathbf{\Phi},\mathbf{\Psi},d+1,T,\textbf{left\_node})$\;
		$\text{GrowFromRoot}(y_{\text{right}},\mathbf{X}_{\text{right}},\mathbf{\Phi},\mathbf{\Psi},d+1,T,\textbf{right\_node})$}
\end{algorithm}
\section{Bayesian Regression Trees for Regression Discontinuity Designs}
\label{sec:xbcf.rdd}
Unlike local polynomial regression methods, a BART-based approach to RDD does not have to pre-specify a set of global basis functions nor must it entirely discard data outside of a neighborhood of the cutoff. These features are particularly useful when incorporating additional covariates $W$ for the purpose of CATE estimation. However, this flexibility comes at a cost and estimation can go wrong in one of two ways. First, a BART-based T-Learner may give poor estimates of $\E(Y \mid X = c, W)$ because tree ensembles with constant leaf models are known to extrapolate poorly (but see \cite{starling2021targeted} and \cite{wang2024local} for alternatives, which we do not pursue here.) Second, a BART-based S-Learner may estimate the {\em response surface} at $X=c$ reasonably well, but still provide biased estimates of the {\em treatment effect} because some of its individual trees end up using data very far from the cutoff. These flaws will be depicted in numerical examples below. 

To overcome these problems, we introduce novel splitting constraints, which ensure that the data used to make predictions at \(X = c\) warrant a causal interpretation. Specifically, we impose the constraint that our ensembles must be composed of trees where any partition containing the point $(x = c,w)$ is estimated from data ``close enough'' to the cutoff from both sides.



\subsection{Splitting Constraints for RDD with Regression Trees}
\label{subsec:constraints}
The proposed constraints have two distinct, though related,
goals. First, we need the treatment-control contrast -- upon
which \(\tau(x=c, w)\) will be estimated -- to be
well-defined: for this we require observations from both
treatment arms (e.g. overlap). Without imposing this
condition it is typical during posterior sampling to
encounter leaf nodes that contain no treated
(resp. untreated) observations, which in turn yields leaf
parameters that are biased for the treatment effect.

Second, because we cannot rely on observations far from the
cutoff to estimate the treatment effect, we insist that a
partition that includes \(x = c\) have a strong majority of
its observations within a narrow, user-defined band about
the cutoff. This constraint defines a set of suitably
\emph{localized} basis functions from which to perform causal
inference at the cutoff.

More formally, these constraints can be expressed as
follows. For a user-defined bandwidth parameter \(h > 0\), we
assume that the potential outcome mean function does not
vary abruptly inside the interval \([c-h,c+h]\), which we
refer to as the ``identification strip''. Let \(B \subset
\mathcal{X}\) be a hypercube corresponding to a node in a
regression tree and let \(N_b\) denote the number of
observations falling within \(B\). Further, let \(n_l\) denote
the number of observations in \(B \cap [c-h,c)\) and \(n_r\)
denote the number of observations in \(B \cap [c,c+h]\). For
user-specified variables \(N_{Omin} \in \mathds{N}^+\) and
\(\alpha \in (0, 1)\), the leaf node region \(B\) is valid if it
satisfies the following condition:
\begin{equation}\label{rdd_constraint}
	\begin{split}
		&\big ( \forall w \mid (x=c, w) \notin B \big) \cup\\
		&\bigg ( \left ( \exists w \mid (x=c, w) \in B \right ) \cap \left
		( \min{(n_{l}, n_{r})} \geq N_{Omin} \right )\cap \left ( (n_l +
		n_r)/N_b \geq \alpha \right) \bigg ).
	\end{split}
\end{equation}
The initial clause says that any node which does not make
predictions at the cutoff remains entirely unrestricted; the
second clause says that any node that \emph{does} make
predictions at \(x = c\) has to have both i) a minimum number
of observations within the cutoff region on either side of
the cutoff, as well as ii) not too many observations,
proportionally, outside of the identification strip.

\subsection{Stochastic search for valid partitions}
For nodes predicting at the cutoff, the two conditions (i
and ii) above have qualitatively different ramifications for
the stochastic search for valid partitions. In particular,
the first condition, if unsatisfied, can never become
satisfied by further branching, while the second condition,
if unsatisfied, \emph{can} be satisfied by further branching, by
trimming away observations outside of the identification
strip. This observation motivates us to use XBART/XBCF
rather than standard BART MCMC for our model fitting. An
unmodified local random walk would violate recurrence
because certain valid states can only be reached by passing
through invalid states; as a practical matter, reaching
valid partitions by a random walk would be highly
inefficient. By utilizing the Grow-From-Root algorithm,
passing through invalid states to reach favorable valid
states is a simple matter of not terminating the growth
process at an invalid state. Specifically, never accept a
partition that violates condition i, and never stop at a
partition that violates condition ii.

In practice, this new stochastic search procedure is
implemented by modifying the likelihood calculation in steps
8 and 10 of the GFR algorithm for XBCF as follows. Consider
a candidate split with cutpoint \(c_{jk}\) which splits the
current node into left and right nodes. Let \(B_x^{(l)}\)
denote the range of \(x\) which the left node covers and
similarly define \(B_x^{(r)}\).  Let \(n_{ll}\) and \(n_{lr}\)
denote the number of observations such that \(x \in [c-h,c)\)
and \(x \in [c,c+h]\) in the left node, and \(n_{rl}\) and
\(n_{rr}\) denote the same quantities in the right node,
respectively. If,
\begin{equation}
	c \in B_x^{(l)} \quad \text{and} \quad \max(n_{ll}, n_{lr}) < N_{Omin} \label{eq:ol.1}
\end{equation}
or if
\begin{equation}
	c \in B_x^{(r)} \quad \text{and} \quad \max(n_{rl}, n_{rr}) < N_{Omin}, \label{eq:ol.2}
\end{equation}
this split violates condition (i). Therefore, we consider
this an invalid partition and set \(L(c_{jk})=0\). If the
split does not violate condition (i), we calculate its
likelihood as in the GFR algorithm. For condition (ii), we
check whether:
\begin{equation}
	c \in B_x \quad \text{and} \quad  \frac{n_l+n_r}{N_b} < \alpha. \label{eq:force.split}
\end{equation}
If so, we set the likelihood of the no-split option
\(L(\emptyset)=0\) unless there are no other valid splits, in
which case we set \(L(\emptyset)=1\). In the latter case, we end up with a tree that is still invalid, as it violates condition ii; our implementation monitors for this eventuality and find that it rarely if ever occurs in most data sets. 

\subsection{Illustration of the constraints and search}
\begin{figure}
	\centering
	\begin{subfigure}[b]{0.45\linewidth}
		\centering
		\includegraphics[width=\linewidth]{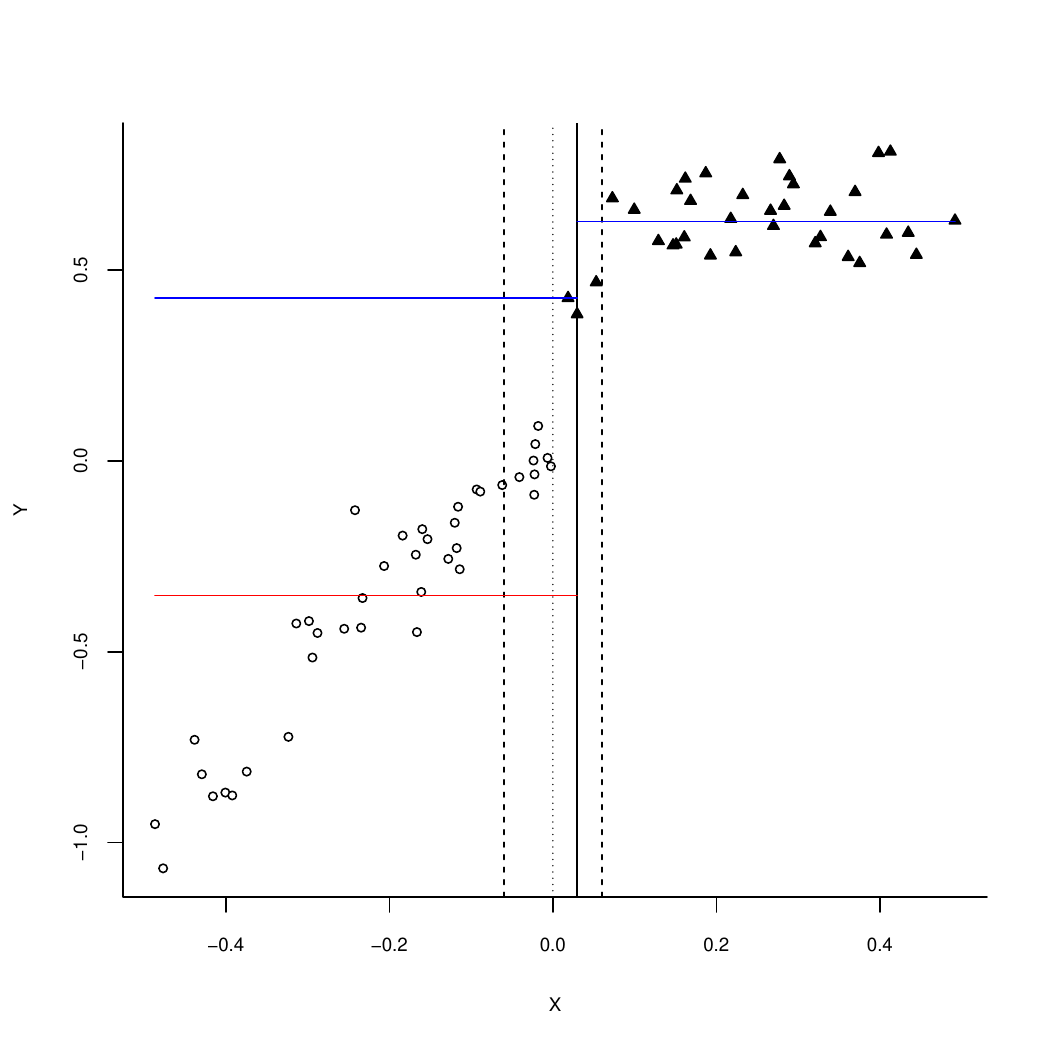}
		\caption{}
		\label{fig:1a}
	\end{subfigure}
	\hfill
	\begin{subfigure}[b]{0.45\linewidth}
		\centering
		\includegraphics[width=\linewidth]{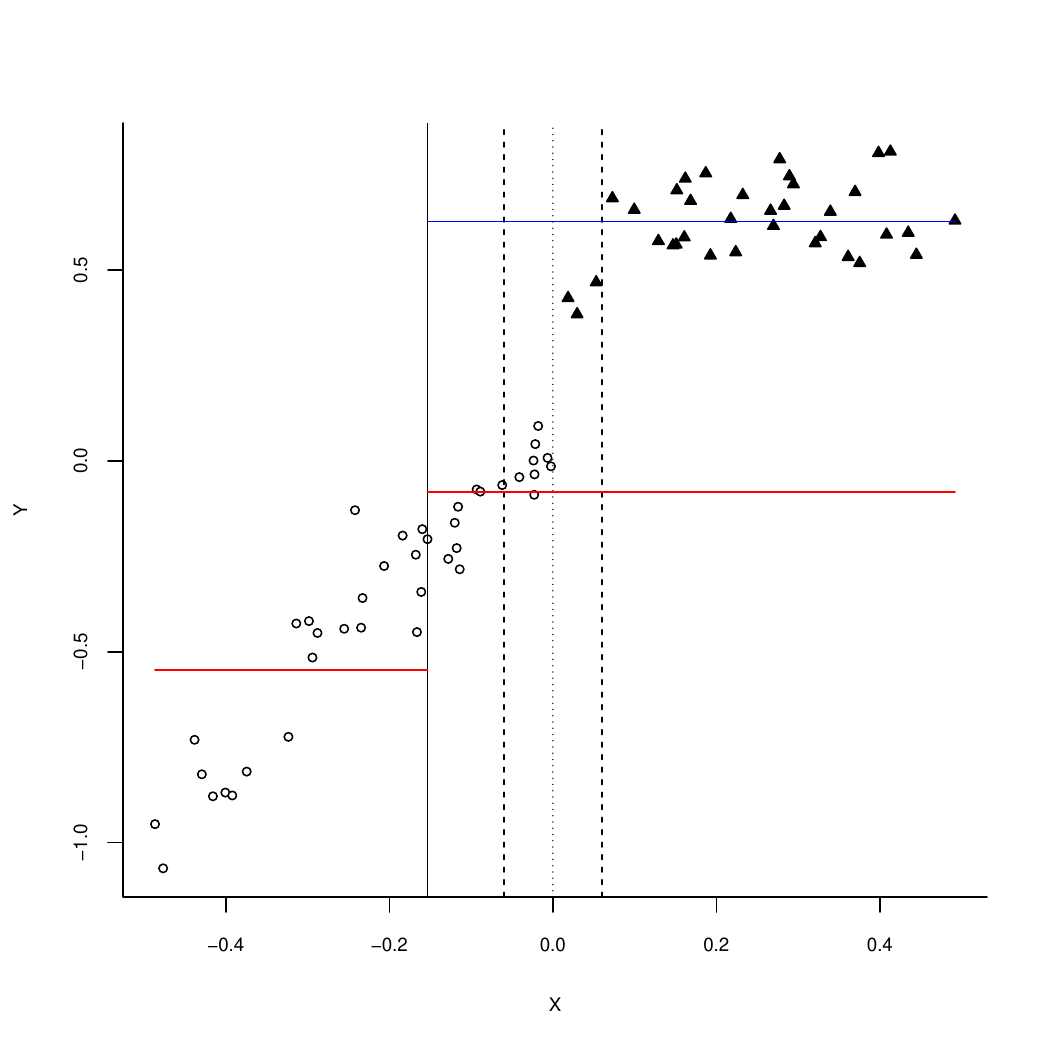}
		\caption{}
		\label{fig:1b}
	\end{subfigure}
	\hfill
	\begin{subfigure}[b]{0.45\linewidth}
		\centering
		\includegraphics[width=\linewidth]{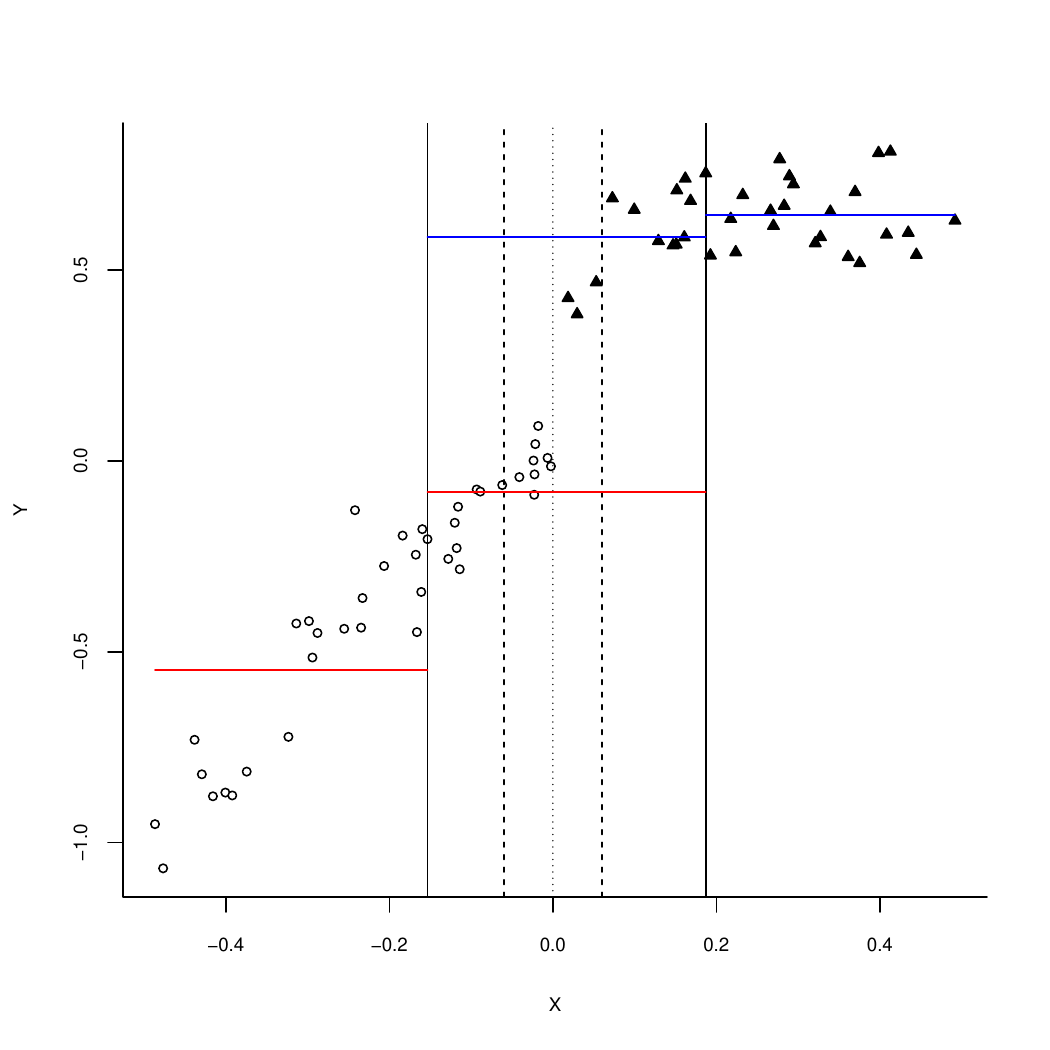}
		\caption{}
		\label{fig:1c}
	\end{subfigure}
	\hfill
	\begin{subfigure}[b]{0.45\linewidth}
		\centering
		\includegraphics[width=\linewidth]{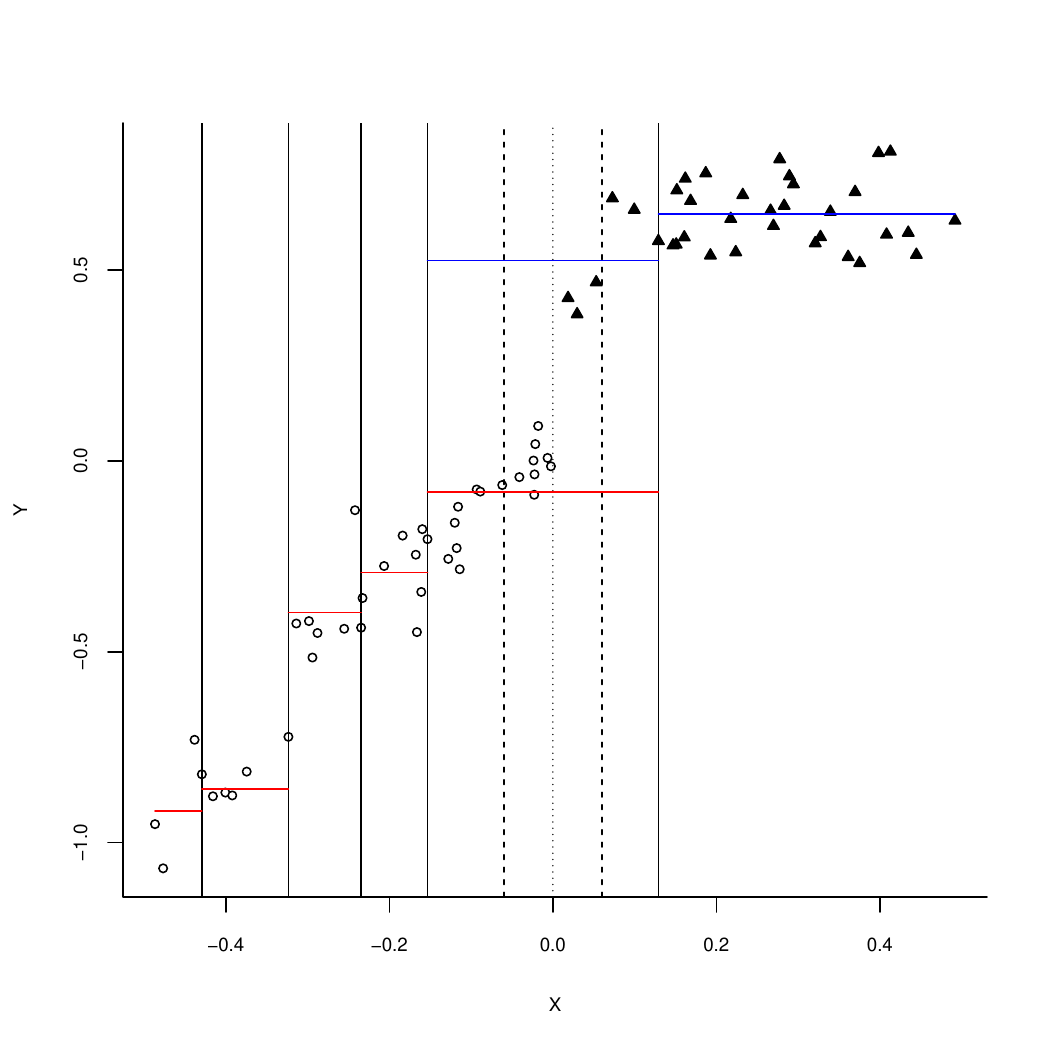}
		\caption{}
		\label{fig:1d}
	\end{subfigure}
	\caption{Tree examples: panel (a) shows a tree with a split that violates condition i, which cannot be accepted; panel (b) presents a tree with a split that violates condition ii, which can be accepted because we can make this tree valid by trimming out the region outside the identification strip in the relevant nodes; the tree in panel (c) is an example of the kind of tree that would be accepted by the algorithm, with tight bounds on the identification region and good representation from both sides of the cutoff; the tree in panel (d) is the same as the one in panel (c) with some additional splits that do not contain the identification strip, so the tree remains valid}
	\label{fig:1d.tree.graphs}
\end{figure}

The impact of expression (\ref{rdd_constraint}) on the fitted trees may be visualized by considering a concrete example. Consider a tree fit with only the running variable ($X$). Figure \ref{fig:1d.tree.graphs} plots $X$ against some outcome $Y$ for a dataset with 75 observations, presenting different partitions in $X$. For this example, the cutoff is $c=0$ -- denoted by the dotted line -- and the ATE at that point is equal to $0.5$. We consider a window of $h=0.06$, denoted by the dashed lines in the plots. The treated units (\(x \geq c\)) are denoted by black triangle dots and the control units are denoted by white round dots. Splits in $X$ are denoted by solid lines. For each partition, we represent the inferred potential outcome as a red line for untreated and blue line for treated outcomes. For the partitions that include both types of observations --- \textit{i.e.} points from both sides of the cutoff --- we represent both potential outcomes.

Panel \ref{fig:1a} shows a split which is invalid since it cuts through the identification strip, leading to a node that contains only one point to the right of the cutoff in that region. The ATE at the cutoff for that tree is predicted to be 0.78. Panel \ref{fig:1b} presents a split which only violates condition (ii), since it does not cut through the identification strip, but features a node with too many points outside the strip. The ATE at the cutoff for that tree is predicted to be 0.7. This tree can be made valid by `trimming out' points too far from the cutoff in the right node. Panel \ref{fig:1c} presents an additional split that does exactly that. The ATE at the cutoff for that tree is predicted to be 0.67. Finally, panel \ref{fig:1d} presents another tree, with a couple of additional splits to the left of the identification strip, and a split to the right that's closer to the strip. Since the new nodes generated do not include the identification strip, they are all potentially valid. The ATE at the cutoff for that tree is predicted to be 0.6.

Analysis of the figures makes clear what types of trees will
be accepted under our restrictions. We consider only trees
that do not cut through the identification strip, are well
populated with points in that region from both sides of the
cutoff and are tight around that region. This way, we
incorporate the RDD assumption that units sufficiently near
the cutoff are similar enough to be compared and use this to
create an `overlap region' around the cutoff. The shape of
the trees is also largely dependent on the data
structure. If there are many points with \(x \approx c\) we
can make the identification strip narrower without being too
restrictive on the tree growth especially if the points are
well dispersed in regards to the other covariates. On the
contrary, if most points have \(x\) far from the cutoff we
might need to define a wider identification strip to
reasonably explore the tree space. Finally, it is worth
noting that this strategy can be used more generally for any
problem where one must fit tree ensembles and enforce
smoothness over a specific variable and around a specific
point.

This exercise also highlights the problems that unmodified
BART models might face in the context of the RDD. Note that
all trees above are, at least in principle, valid under the
standard BART prior. If the nodes that contain \(X=c\) include
points close to the cutoff from both sides, but many points
far from it, these trees will only lead to reasonable causal
contrasts if \(Y\) is relatively constant with respect to \(X\). Otherwise, such trees should exhibit strong bias if the prognostic or treatment effect functions vary substantially, as is the case in the previous example, which illustrates the bias described in equation \eqref{eq:bias}. In fact, as we move closer towards the kinds of trees that would be accepted by BART-RDD --- \textit{i.e.} moving from the first panel to the last --- we decrease the bias in the predicted ATE at the cutoff. The important distinction here is
that, while all BART-based models could reach reasonable
trees, only BART-RDD is guaranteed to
do so by rejecting trees that do not behave `well'. This means the unmodified BART trees might sometimes do well and sometimes not, so these models will produce ensembles which mix over biased and unbiased trees, leading to biased fits.
\subsection{Prior elicitation}
\label{sec:prior.elicitation}
The question remains as to how one should set the relevant
parameters in order to obtain good predictions. A completely general rule cannot be expected, as the impact each
parameter has in the estimation is highly
data-dependent. Nonetheless, it is instructive to
consider what kinds of restrictions our parameters imply
to the tree search process and how they impact prior bias.

First, consider the bandwidth parameter $h$. On the one hand, \(h\) should be set as low as possible for two
reasons. First, setting \(h\) too high makes it more likely
that a given tree would cut through the identification
strip. In particular if \(h\) is such that the strip covers
all the support of \(X\), the algorithm would only accept
trees that do not use \(X\) for partitions. Given the
essential role that \(X\) plays in the outcome distribution in
an RDD setting, not using it for the tree splits would lead
to severely biased trees. Additionally, since the goal is to
make inference enforcing smoothness over \(X\) at a specific
point (\(X=c\)), one should only use points as close to \(c\) as
possible to obtain better approximations of the true
function around that point.

On the other hand, there is also a limit to how low one can
set \(h\) for each dataset. In particular, if \(h\) is so small
that there are no points inside the identification strip,
any tree will produce nodes with an empty overlap region
and, thus, be invalid. This means that \(h\) also interacts
with \(N_{Omin}\) in that extreme: even if there
are points in the identification strip, if there are less
than \(N_{Omin}\) points, the same phenomenon happens, making
all trees invalid.

Next, we turn to \(N_{Omin}\): if it is set to 0, the trees could
produce nodes which contain the overlap region but have no
points inside it. Thus, predictions near the cutoff could be
based only on observations too far from the cutoff, which
would undermine the constraint associated with this
parameter. Setting \(N_{Omin}\) too high could be too
restrictive since there would be fewer valid nodes containing
the overlap region, which could bias the individual
posterior distributions or at least make it harder to detect
heterogeneity in the data, since we could have very short
trees.

Finally, we note that if \(\alpha\) is set too low, we allow for many
points outside the strip to influence the results from nodes
that do include the strip. However, setting it high is not
necessarily a problem since, as discussed in the previous
section, nodes that do not satisfy this criterion can be
made to satisfy it by simply `trimming' the outer region of
nodes containing the identification strip. It is important to note, however, that setting $\alpha$ too high could lead to many forced splits in the boundaries of the identification strip, which can lead to an increase in variance if these additional splits are not particularly relevant. Therefore, this parameter should also be chosen carefully.

Clearly, it is nontrivial how these considerations might
interact in a given sample. This reflects the immense delicacy of the regression discontinuity design itself, rather than a limitation intrinsic to the BART-RDD proposed; all RDD methods require grappling with how to set tuning parameters. Our proposed approach is via a {\em prior predictive elicitation} procedure. Specifically, we recommend, for a given
sample \((x,w,y)\), the following:

\begin{enumerate}
	\item Generate \(s\) samples of a synthetic data from a known DGP
	using \((x,w)\)
	\item Fit BART-RDD to each sample for different values of the prior parameters
	\item Choose the parameter values which lead to lower RMSE values for the ATE in those \(s\) synthetic samples.
\end{enumerate}

We find that generating the synthetic data from a simple
model with no treatment effect heterogeneity and relatively small ATE leads to finding good values for the prior parameters even when the true data exhibits strong heterogeneity or large effects. This is also a reasonable prior for the treatment effects, unless one has strong reason to believe in a more complex scenario. In the simulation studies to follow, we use this procedure to choose $h$, $N_{Omin}$ and $\alpha$. A detailed analysis of this procedure may be found in Appendix \ref{calibration}, where we discuss its application in the context of the simulations.
\section{Simulation studies}
\label{sec:simulations}
\subsection{Setup}
\label{sec:org35deff4}
In order to investigate the properties of the BART-RDD
algorithm, we perform a simulation study comparing its
performance to an S-learner BART fit (S-BART), a T-learner BART fit (T-BART), the robust bias-corrected
local linear regression (LLR), as implemented by
\citet{calonico2015rdrobust}, and the cubic spline estimator
(CGS) of \citet{chib2023nonparametric}. The goal of this
exercise is twofold. First, we want to investigate how
BART-RDD compares with off-the-shelf implementations of BART
applied to the RDD context. We are able to show that our
modification does in fact make the BART prior more suited to
this context. Second, we want to compare BART-RDD to
estimators that were designed specifically to the RDD
context, in particular the local polynomial estimator, by
far the most commonly used in the literature, and the cubic
splines estimator which is possibly the closest in spirit to
that in the Bayesian literature. Besides showing BART-RDD is
more suited to the RDD setup than unmodified BART models, we
also show that BART-RDD generally performs better and never far worse than the estimators designed specifically for the RDD.

Let \(X\) denote the running variable, \(W\) an additional set
of features, \(Z\) the treatment indicator and \(Y\) a
continuous outcome. We investigate 500 samples of size 1000
of variations of the following DGP:\\
\begin{minipage}{0.3\linewidth}
	\begin{align*}
			X &\sim 2 \mathcal{B}(2,4)-0.75\\
			W_1 &\sim U(-0.1,0.1)\\
			W_2 &\sim \mathcal{N} (0,0.2)\\
			W_3 &\sim \text{Binomial}(1,0.4)\\
			W_4 &\sim \text{Binomial}(1,p(x))
	\end{align*}
\end{minipage}%
\begin{minipage}{0.3\linewidth}
	\begin{align*}
		&Z = \mathbf{1}(X \geq c)\\
		&\mu(X,W) = \frac{\mu_0(X,W)}{\sigma(\mu_0(X,W))} \delta_{\mu}\\
		&\tau(X,W) = \bar{\tau} + \frac{\tau_0(X,W)}{\sigma(\tau_0(X,W))} \delta_{\tau}\\
		&Y = \mu(X,W) + \tau(X,W) Z + \varepsilon
	\end{align*}
\end{minipage}%
\begin{minipage}{0.3\linewidth}
	\begin{align*}
		c &= 0\\
		\bar{\tau} &= \{0.2,0.5\}\\
		\delta_{\mu} &= \{0.5,1.25\}\\
		\delta_{\tau} &= \{0.1,0.3\}\\
		\varepsilon &\sim \mathcal{N} (0,1),
	\end{align*}
\end{minipage}\\[5pt]
where \(\mathcal{B}(2,4)\) denotes a Beta distribution with
parameters 2 and 4, $p(x)$ denotes the Gaussian probability density of $x$ with mean $c$ and standard deviation $0.5$, and:

\begin{equation}
	\begin{split}
		\mu_0(X,W) &= 3 x^5 - 2.5 x^4 - 1.5 x^3 + 2 x^2 + 3 x + 2 + \frac{1}{2} \sum_{p=1}^4 (w_p-E[w_p])\\ \tau_0(X,W) &= - 0.1 x + \frac{1}{4} \sum_{p=1}^4 (w_p-E[w_p])
	\end{split}
\end{equation}

\subsubsection{Rationale}
\label{sec:org7156957}
Here we briefly justify the choices made in the simulation design
described above. First, although there are other
parameters that affect the performance of any estimator, the spread in $\mu$ and $\tau$ were the only factors that we found to have distinct impacts on different estimators. In other words, the effect of other DGP characteristics in the results were common across estimators in the expected ways\footnote{For example, larger sample or effect sizes increased the performance of every estimator in roughly the same manner, meaning these features are not particularly helpful in determining the situations in which the estimators might differ in their performance.}. We control these features in the data through the parameters $(\delta_{\mu},\delta_{\tau})$. The particular choices for these parameters were made in an attempt to replicate realistic behavior in $\mu$ and $\tau$. Particularly, we made sure that there are generally no sign changes in the individual treatment effects and that the spread in the prognostic component is larger than the spread in the treatment effects.

In regards to the functional forms chosen, while we did experiment with different functional
forms, the results remain the same qualitatively
(although sometimes less clear depending on how hard the
functions are to estimate). In most methodological RDD studies, the setups
considered feature only \(X\) as a strong predictor and a very
strong signal-to-noise ratio. In that regard, we consider
our setup to be more complete in terms of expected characteristics of
empirical data\footnote{For a summary of the simulation exercises in some of the most relevant methodological RDD papres, see \url{https://github.com/rafaelcalcantara/BART-RDD}.}.

The distribution of \(X\) plays an important role in the RDD,
as it determines the distribution of treatment. If \(X\) is
skewed to the left (right) of the cutoff, the sample will
have many less (more) treated units, which should make
estimation harder. Conversely, if \(X\) is nearly symmetric
around the cutoff, estimation should be simpler. The
distribution described above is relatively standard in the
literature, so we chose it for more comparability with
previous studies\footnote{Most papers in fact set \(X \sim 2
	\mathcal{B}(2,4)-1\). We chose \(X \sim 2
	\mathcal{B}(2,4)-0.75\) here so that \(X\) is centered slightly
	closer to the cutoff and the proportion of treated and
	control units in the sample is not so different (we obtain
	nearly \(40\%\) treated units in every sample).}. While we did explore different
distributions of \(X\), these variations did not
change the results qualitatively.

\subsection{Results}
\subsubsection{Comparison of ATE Estimates}
\begin{figure}
	\centering
	\begin{subfigure}[b]{0.49\textwidth}
		\centering
		\includegraphics[width=\textwidth]{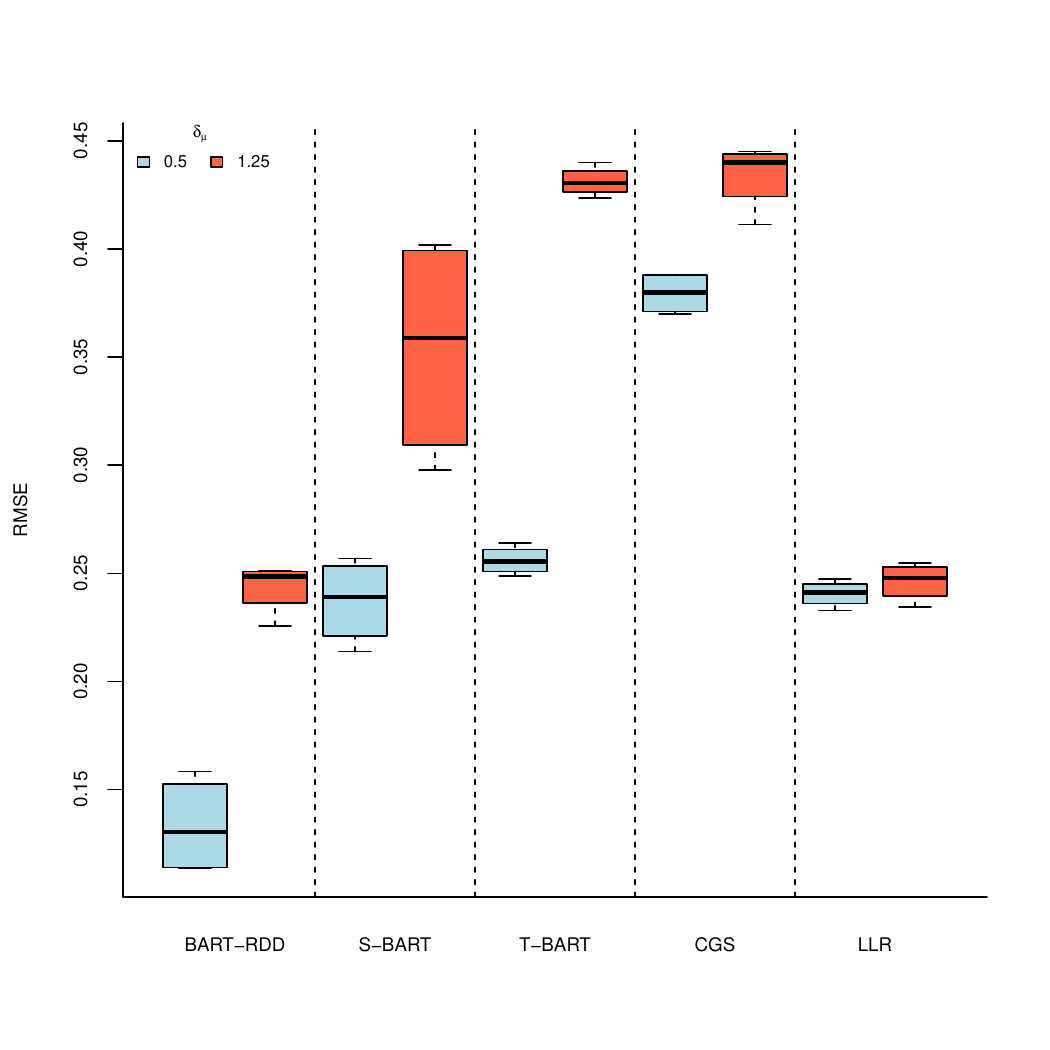}
		\caption{$\delta_{\mu}$}
		\label{fig:rmse.delta.mu}
	\end{subfigure}
	\hfill
	\begin{subfigure}[b]{0.49\textwidth}
		\centering
		\includegraphics[width=\textwidth]{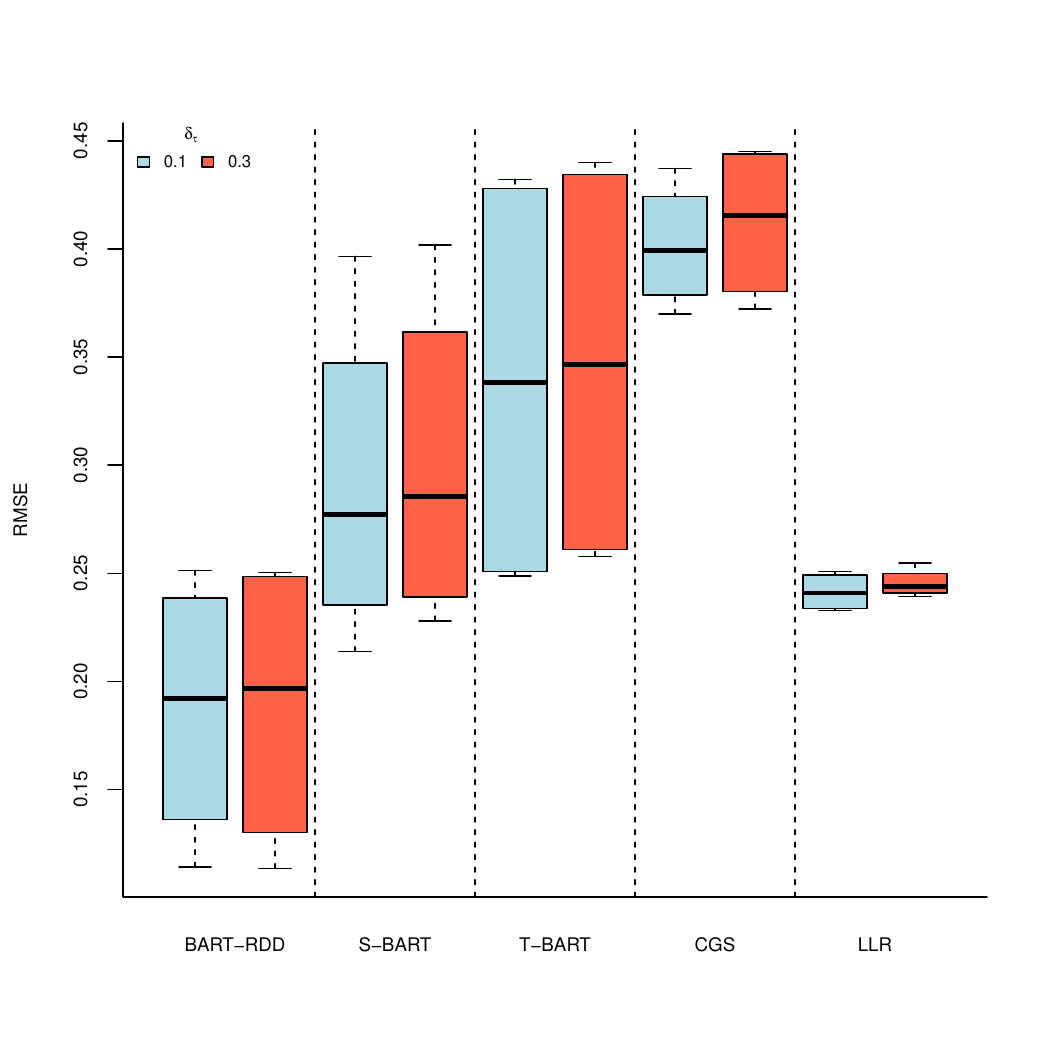}
		\caption{$\delta_{\tau}$}
		\label{fig:rmse.delta.tau}
	\end{subfigure}
	\hfill
	\begin{subfigure}[b]{0.49\textwidth}
		\centering
		\includegraphics[width=\textwidth]{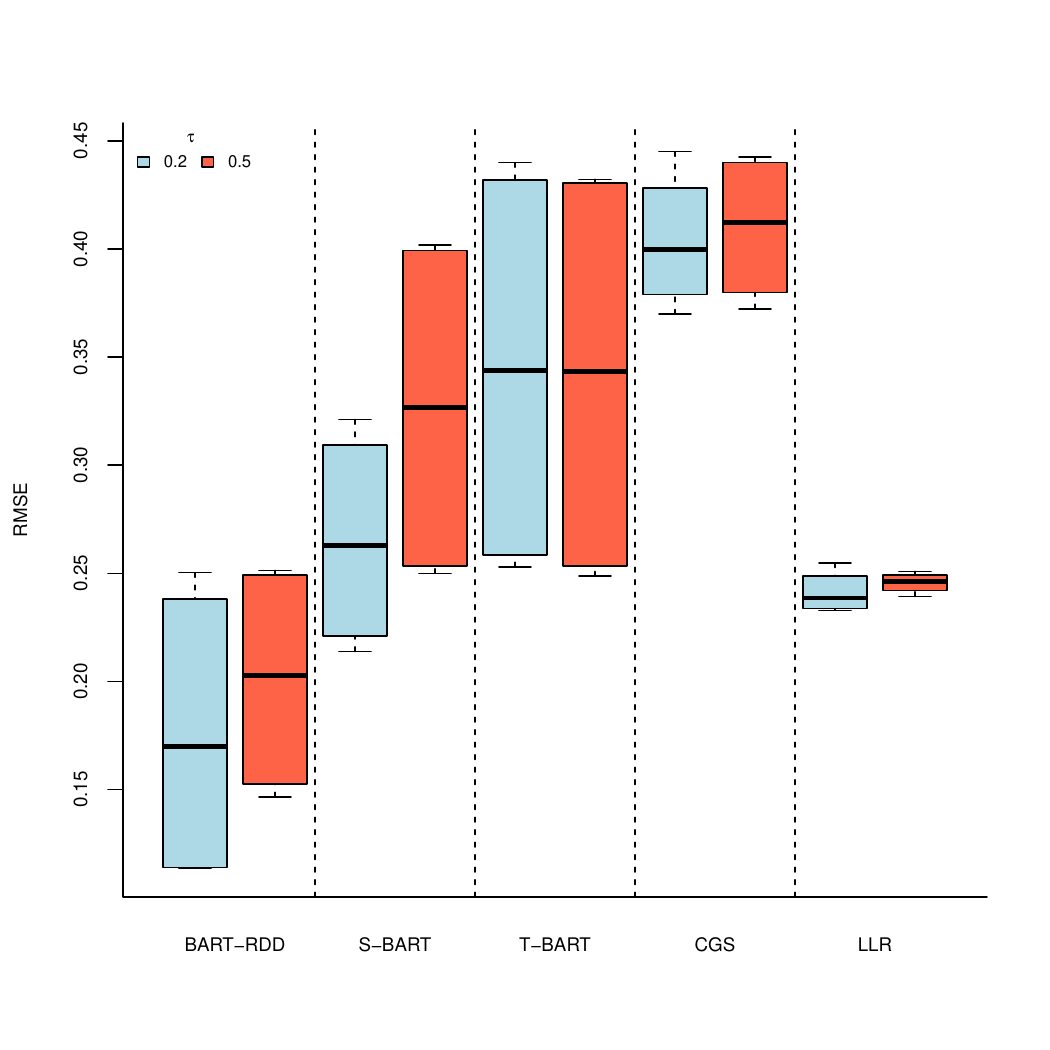}
		\caption{$\tau$}
		\label{fig:rmse.tau}
	\end{subfigure}
	\caption{RMSE for the ATE point estimate produced by each estimator, divided by the different DGP parameters}
	\label{fig:sim.rmse.ate}
\end{figure}

Although the primary new functionality of BART-RDD is in providing CATE estimates, we begin by examining its performance on the ATE for comparison with other methods and because a good CATE learner should be able to provide good ATE estimates as well. 
\label{sec:org3138993}
Table \ref{tab:ate.rmse} and figure \ref{fig:sim.rmse.ate} present the RMSE for the ATE point estimate produced by each estimator\footnote{In the case of the Bayesian estimators, we consider the posterior mean as the point estimate.}.

	\begin{table}[!htbp] \centering 
		\caption{RMSE - ATE} 
		\label{tab:ate.rmse} 
		\small
		\begin{tabular}{@{\extracolsep{5pt}} cccccccc}  
			\hline \\[-1.8ex] 
			$\bar{\tau}$ & $\delta_{\mu}$ & $\delta_{\tau}$ & BART-RDD & S-BART & T-BART & CGS & LLR \\ 
			\hline \\[-1.8ex] 
			$0.2$ & $0.5$ & $0.1$ & $0.114$ & $0.214$ & $0.253$ & $0.370$ & $0.233$ \\ 
			$0.2$ & $0.5$ & $0.3$ & $0.114$ & $0.228$ & $0.264$ & $0.388$ & $0.243$ \\ 
			$0.2$ & $1.25$ & $0.1$ & $0.226$ & $0.298$ & $0.424$ & $0.411$ & $0.234$ \\ 
			$0.2$ & $1.25$ & $0.3$ & $0.250$ & $0.321$ & $0.440$ & $0.445$ & $0.255$ \\ 
			$0.5$ & $0.5$ & $0.1$ & $0.158$ & $0.257$ & $0.249$ & $0.387$ & $0.247$ \\ 
			$0.5$ & $0.5$ & $0.3$ & $0.147$ & $0.250$ & $0.258$ & $0.372$ & $0.239$ \\ 
			$0.5$ & $1.25$ & $0.1$ & $0.251$ & $0.397$ & $0.432$ & $0.437$ & $0.251$ \\ 
			$0.5$ & $1.25$ & $0.3$ & $0.247$ & $0.402$ & $0.429$ & $0.443$ & $0.245$ \\
			\hline \\[-1.8ex] 
		\end{tabular} 
	\end{table} 

The results indicate that high variation in $\mu$ makes estimation harder for all estimators, although the difference is not so sizeable for LLR. In that setting, BART-RDD and LLR perform similarly. However, when $\delta_{\mu}$ is low, BART-RDD clearly outperforms all estimators. Regarding the other BART-based estimators, S-BART and T-BART perform similarly, but the former is less sensitive to high variability in $\mu$. Overall, CGS is the worst performer in terms of the RMSE.

\begin{figure}
	\centering
	\begin{subfigure}[b]{0.49\textwidth}
		\centering
		\includegraphics[width=\textwidth]{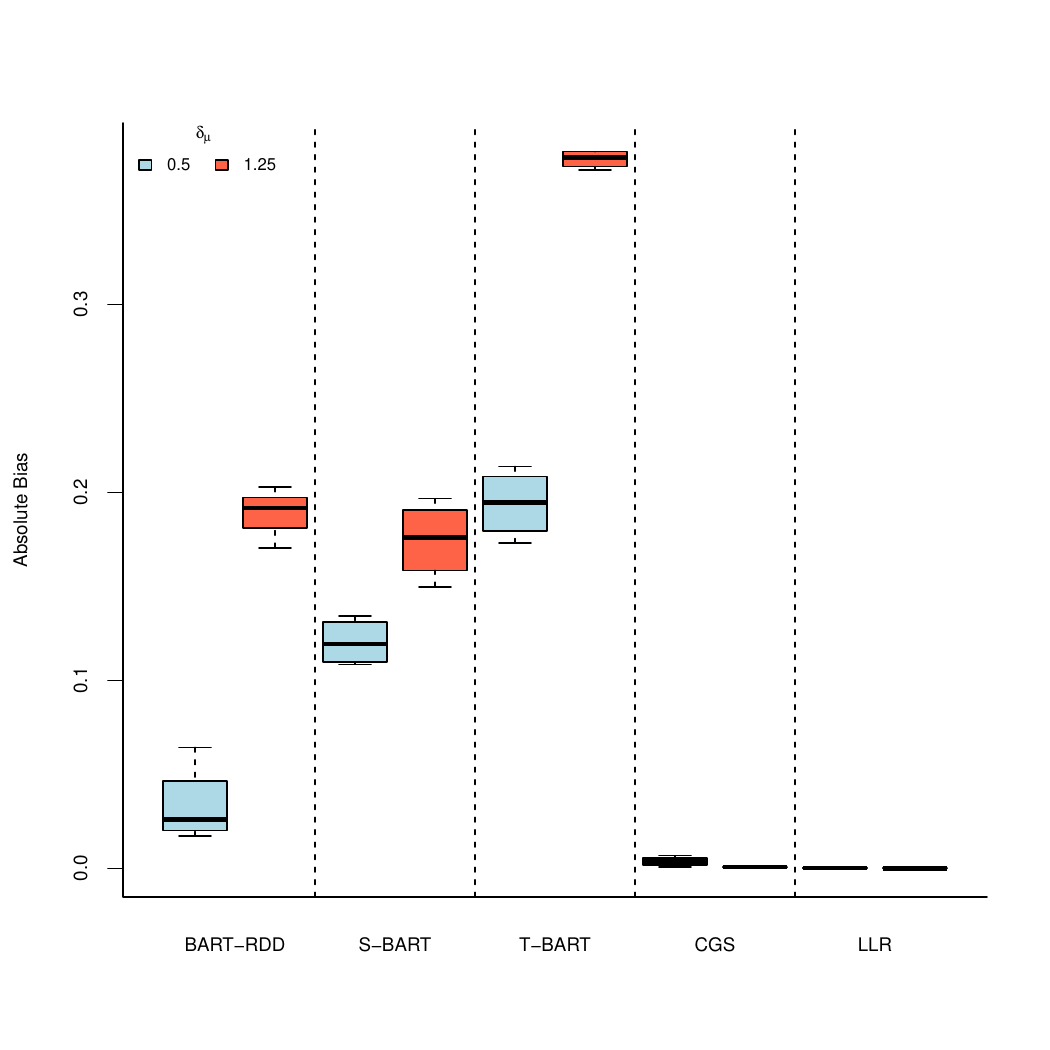}
		\caption{$\delta_{\mu}$}
		\label{fig:bias.delta.mu}
	\end{subfigure}
	\hfill
	\begin{subfigure}[b]{0.49\textwidth}
		\centering
		\includegraphics[width=\textwidth]{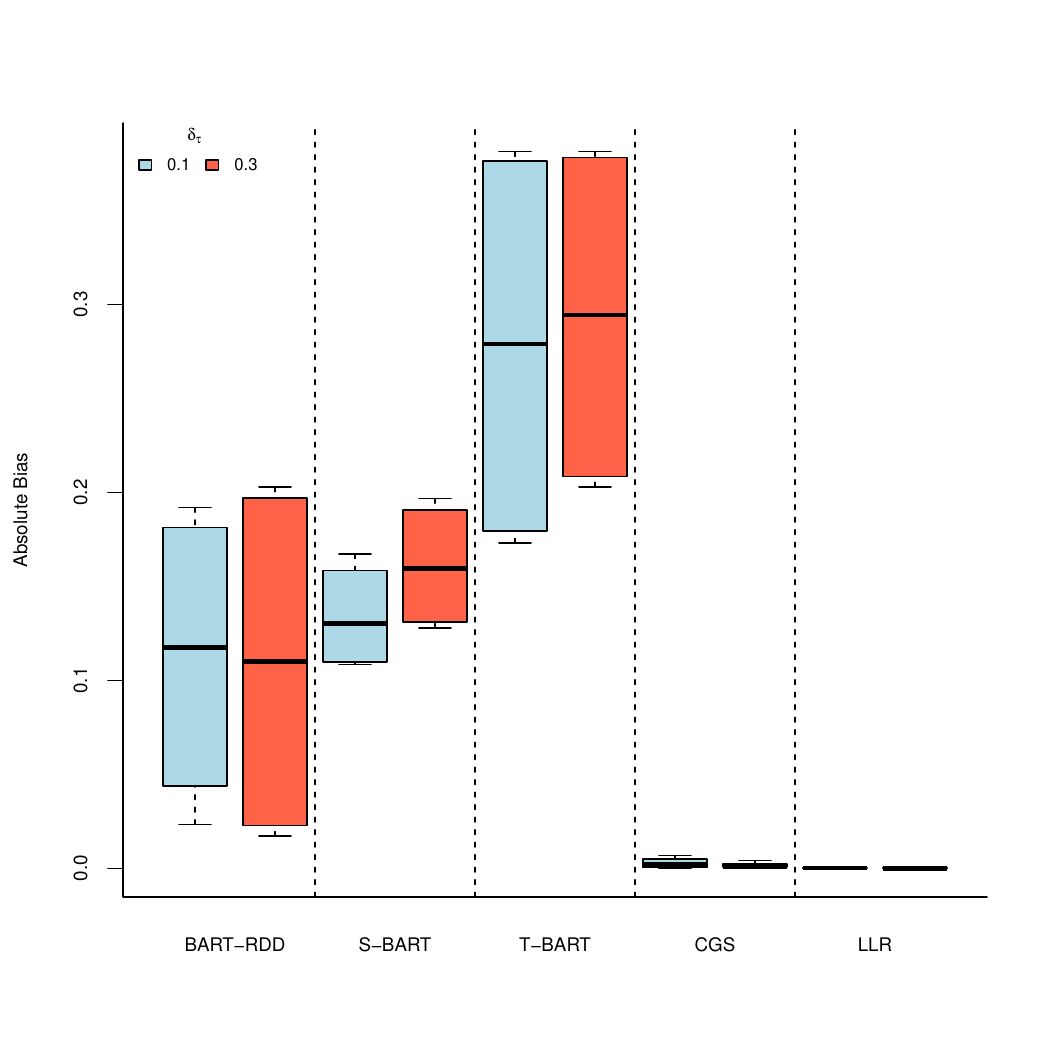}
		\caption{$\delta_{\tau}$}
		\label{fig:bias.delta.tau}
	\end{subfigure}
	\hfill
	\begin{subfigure}[b]{0.49\textwidth}
		\centering
		\includegraphics[width=\textwidth]{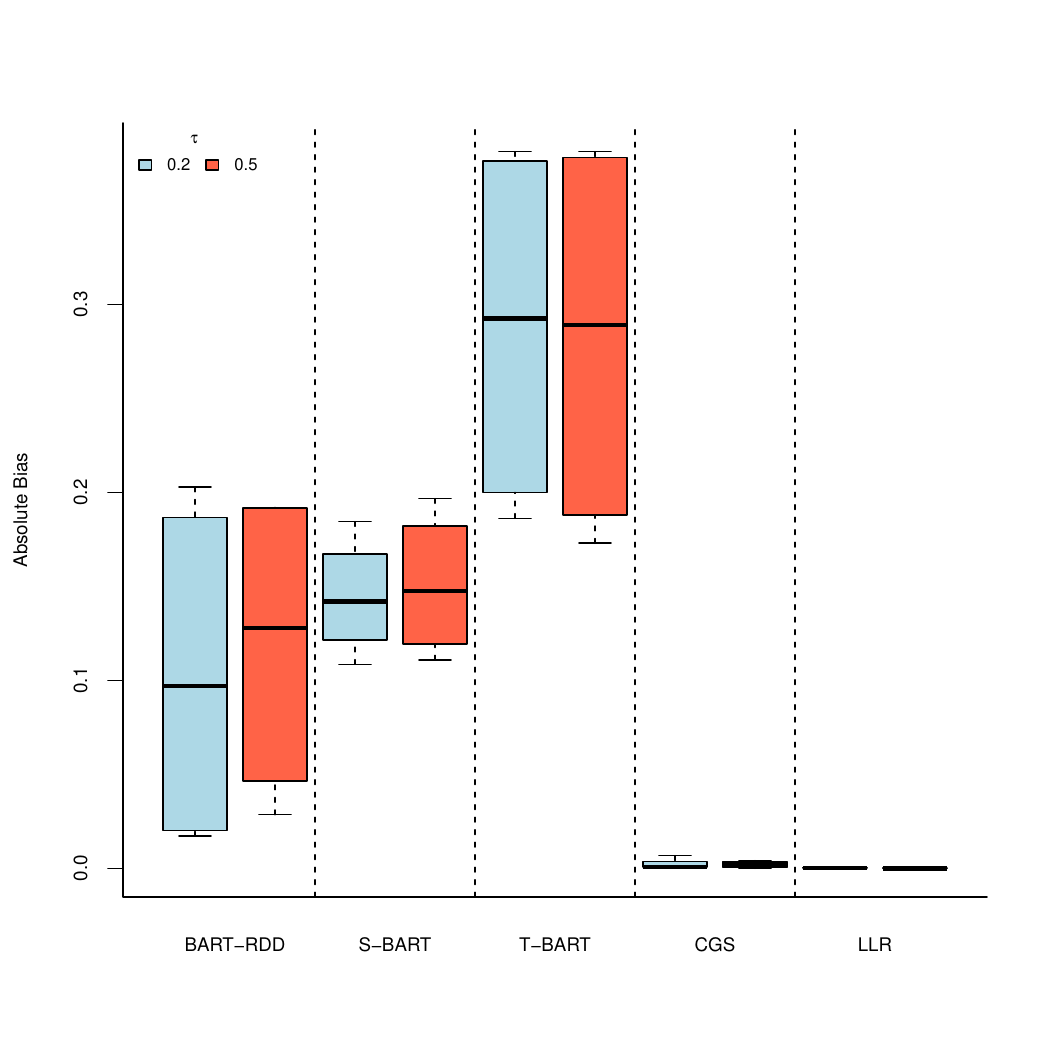}
		\caption{$\tau$}
		\label{fig:bias.tau}
	\end{subfigure}
	\caption{Absolute bias for the ATE point estimate produced by each estimator, divided by the different DGP parameters}
	\label{fig:sim.bias.ate}
\end{figure}

\begin{figure}
	\centering
	\begin{subfigure}[b]{0.49\textwidth}
		\centering
		\includegraphics[width=\textwidth]{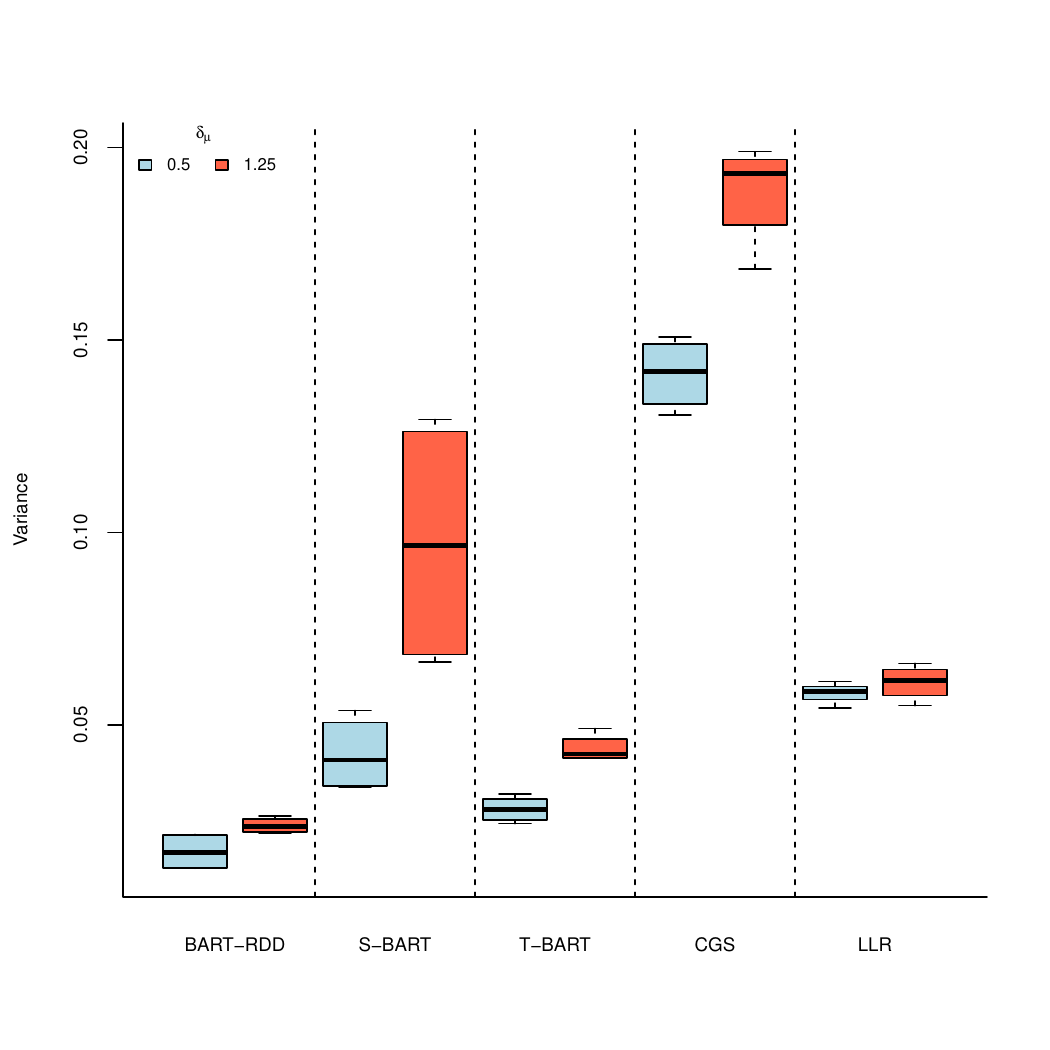}
		\caption{$\delta_{\mu}$}
		\label{fig:var.delta.mu}
	\end{subfigure}
	\hfill
	\begin{subfigure}[b]{0.49\textwidth}
		\centering
		\includegraphics[width=\textwidth]{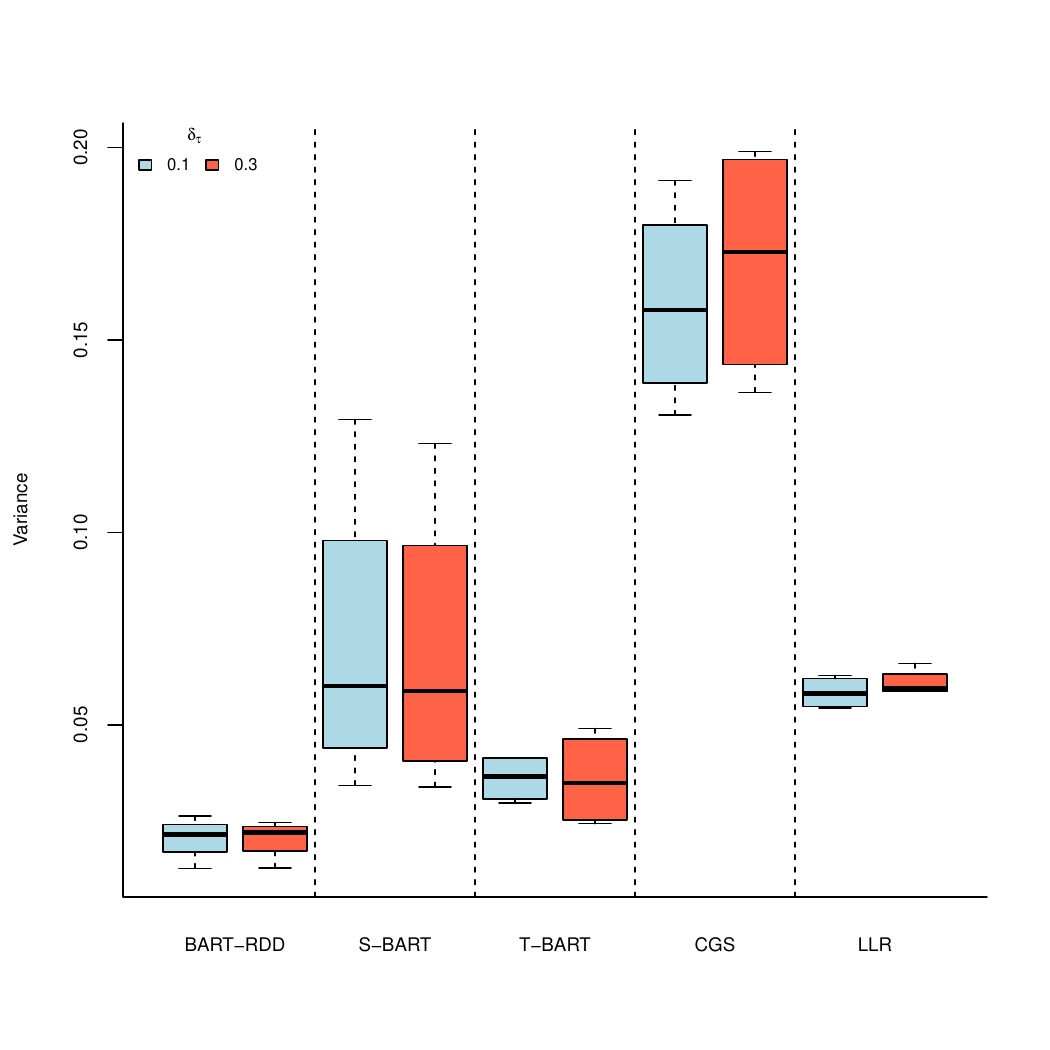}
		\caption{$\delta_{\tau}$}
		\label{fig:var.delta.tau}
	\end{subfigure}
	\hfill
	\begin{subfigure}[b]{0.49\textwidth}
		\centering
		\includegraphics[width=\textwidth]{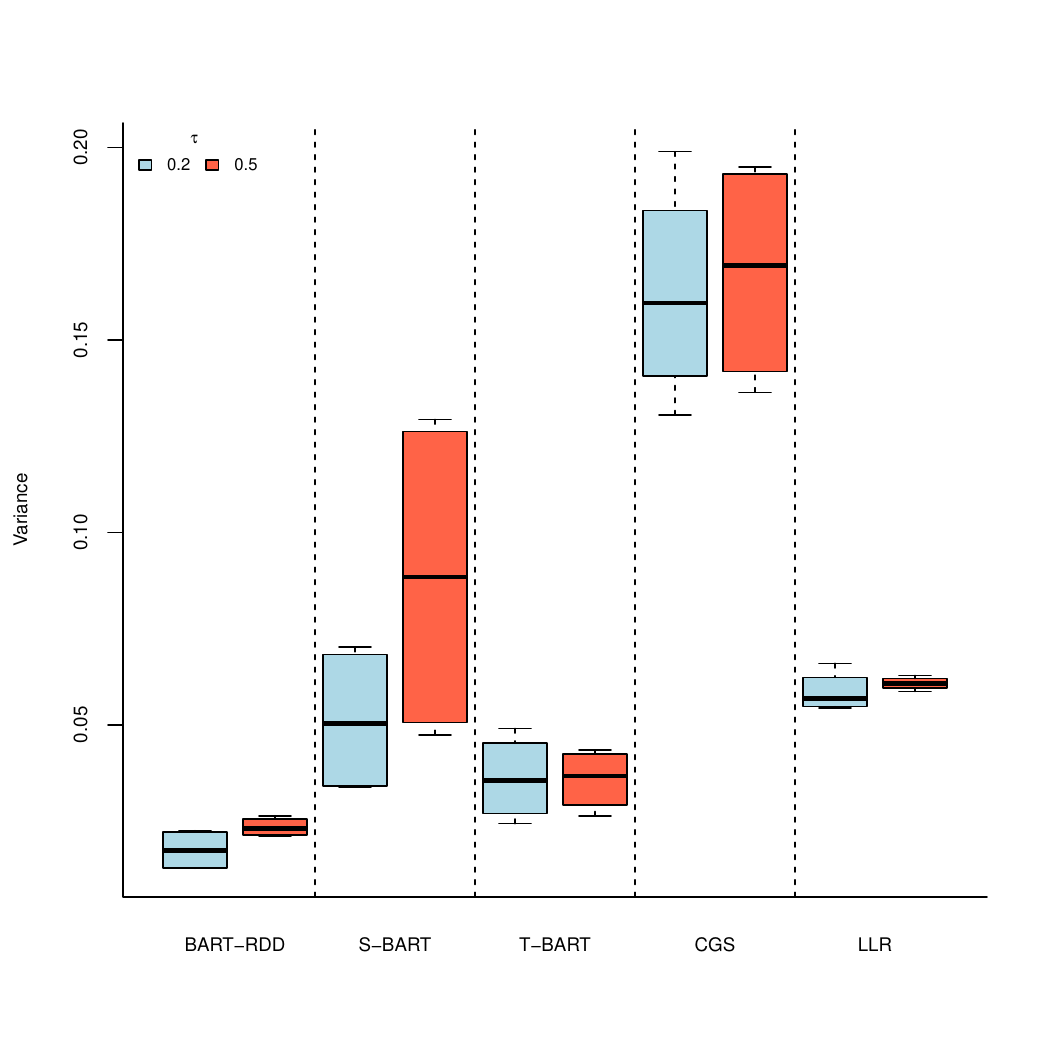}
		\caption{$\tau$}
		\label{fig:var.tau}
	\end{subfigure}
	\caption{Variance for the ATE point estimate produced by each estimator, divided by the different DGP parameters}
	\label{fig:sim.var.ate}
\end{figure}

In order to better understand the behavior of the RMSE,
figures \ref{fig:sim.bias.ate} and \ref{fig:sim.var.ate} present, respectively, the absolute
bias and variance for each estimator, separated by the parameter values of the DGPs. This decomposition highlights
some important patterns. First, the consistently low bias of the LLR and CGS estimators is remarkable, which
means any variation in their RMSE is coming from the estimator variance. For LLR, this should not come as a surprise given this method's focus on reducing bias, but it is worth noting how effective it is in that regard. On the contrary,
BART-RDD presents bias comparable to LLR and CGS when heterogeneity in $\mu$ is low, but a much greater bias otherwise. This trend is true for all BART-based models, although, for a given value of $\delta_{\mu}$, BART-RDD almost always presents much lower --- and never far worse --- bias than the others. Finally, $\delta_{\mu}$ is the only factor that significantly affects bias for the BART-based models. These results corroborate the bias described in equation \eqref{eq:bias} for tree-based RDD estimators. In particular, although both variation in $\mu$ and $\tau$ near the cutoff point can pose problems, the models are potentially much more sensitive to the former, since they require low variation for the prognostic function at both sides of the cutoff. The results also highlight how BART-RDD is particularly effective in decreasing the off-the-shelf BART sensititivity to such issues

Regarding variance, BART-RDD is always the best performer, with a consistently lower variance than the other estimators. LLR presents much larger variance than BART-RDD. T-BART presents a slightly larger variance than BART-RDD, whereas S-BART presents larger variance that is very sensitive to $\delta_{\mu}$. Finally, CGS presents the worst variance in all scenarios, which explains this method's poor RMSE performance.

Although Bayesian intervals are generally not expected to achieve any particular coverage rate, frequentist coverage is a helpful metric to consider. Table \ref{tab:cov.ci} presents the coverage rate and interval size (in parenthesis) for each estimator. CGS and LLR present near 95\% coverage in all cases, while S-BART presents near 95\% coverage in all cases when $\bar{\tau}=0.2$ and near 90\% coverage when $\bar{\tau}=0.5$. Coverage rates for BART-RDD and T-BART follow a common pattern of decreasing coverage when $\delta_{\mu}$ increases. However BART-RDD presents better coverage than T-BART, reaching 95\% coverage in one case and never falling below 70\% coverage. Meanwhile, T-BART reaches at most 82.8\% coverage.

Comparing the interval sizes provides an explanation of the coverage rate behavior. CGS presents, by far, the largest intervals so it is still able to get good coverage despite being the worst in terms of the RMSE. S-BART produces the second largest intervals on average, which also helps compensate the larger bias in some cases, leading to very good coverage generally. LLR produces the third largest intervals on average, which, combined with the relatively good RMSE performance leads to great coverage. T-BART comes next, and the combination of shorter intervals and bad RMSE performance leads to poor coverage. Finally, BART-RDD produces the shortest intervals. However, because of the really good RMSE performance, it is still able to obtain good coverage in all cases.

\begin{table}[!htbp] \centering 
  \caption{Coverage rate and interval sizes (in parenthesis) for the ATE} 
  \label{tab:cov.ci} 
\small 
\begin{tabular}{@{\extracolsep{0pt}} cccccccc} 
\hline 
$\tau$ & $\delta_{\mu}$ & $\delta_{\tau}$ & BART-RDD & S-BART & T-BART & CGS & LLR \\ 
\hline
0.2 & 0.5 & 0.1 & 0.924 & 0.954 & 0.798 & 0.962 & 0.94 \\ 
 &  &  & (0.424) & (0.713) & (0.719) & (1.598) & (0.855) \\ 
0.2 & 0.5 & 0.3 & 0.954 & 0.95 & 0.724 & 0.95 & 0.932 \\ 
 &  &  & (0.442) & (0.757) & (0.677) & (1.604) & (0.877) \\ 
0.2 & 1.25 & 0.1 & 0.782 & 0.94 & 0.538 & 0.97 & 0.93 \\ 
 &  &  & (0.546) & (0.97) & (0.797) & (1.792) & (0.863) \\ 
0.2 & 1.25 & 0.3 & 0.718 & 0.95 & 0.52 & 0.964 & 0.938 \\ 
 &  &  & (0.539) & (1.068) & (0.794) & (1.814) & (0.88) \\ 
0.5 & 0.5 & 0.1 & 0.9 & 0.866 & 0.828 & 0.958 & 0.946 \\ 
 &  &  & (0.536) & (0.859) & (0.743) & (1.604) & (0.87) \\ 
0.5 & 0.5 & 0.3 & 0.92 & 0.89 & 0.772 & 0.962 & 0.942 \\ 
 &  &  & (0.519) & (0.913) & (0.704) & (1.607) & (0.87) \\ 
0.5 & 1.25 & 0.1 & 0.722 & 0.87 & 0.558 & 0.966 & 0.918 \\ 
 &  &  & (0.579) & (1.239) & (0.81) & (1.788) & (0.87) \\ 
0.5 & 1.25 & 0.3 & 0.702 & 0.894 & 0.572 & 0.962 & 0.934 \\ 
 &  &  & (0.567) & (1.313) & (0.818) & (1.798) & (0.872) \\ 
\hline
\end{tabular} 
\end{table}

\subsubsection{Comparison of CATE estimates}
This section compares the various BART-based models in terms of their CATE estimation (the polynomial estimators do not provide CATE estimates). Tables \ref{tab:cate.rmse} and \ref{tab:coverage.cate} present the RMSE and coverage for each estimator, respectively. The results for the RMSE are qualitatively the same as before for all estimators. Regarding coverage, BART-RDD is the best model, with S-BART and T-BART performing slightly worse. Overall, these results suggest a similar trend as with the ATE: S-BART and T-BART present similar performance, with the latter being more sensitive to variability in $\mu$. BART-RDD comes out as the best estimator among the BART variants in all scenarios but one.

\begin{table}[!htbp] \centering 
	\caption{RMSE - CATE} 
	\label{tab:cate.rmse} 
	\small
	\begin{tabular}{@{\extracolsep{5pt}} cccccc} 
		\hline
		$\tau$ & $\delta_{\mu}$ & $\delta_{\tau}$ & BART-RDD & S-BART & T-BART \\ 
		\hline \\[-1.8ex] 
		$0.2$ & $0.5$ & $0.1$ & $0.164$ & $0.204$ & $0.280$ \\ 
		$0.2$ & $0.5$ & $0.3$ & $0.216$ & $0.287$ & $0.298$ \\ 
		$0.2$ & $1.25$ & $0.1$ & $0.262$ & $0.255$ & $0.445$ \\ 
		$0.2$ & $1.25$ & $0.3$ & $0.302$ & $0.345$ & $0.463$ \\ 
		$0.5$ & $0.5$ & $0.1$ & $0.228$ & $0.247$ & $0.281$ \\ 
		$0.5$ & $0.5$ & $0.3$ & $0.249$ & $0.297$ & $0.295$ \\ 
		$0.5$ & $1.25$ & $0.1$ & $0.315$ & $0.363$ & $0.451$ \\ 
		$0.5$ & $1.25$ & $0.3$ & $0.321$ & $0.411$ & $0.452$ \\ 
		\hline \\[-1.8ex] 
	\end{tabular} 
\end{table} 

\begin{table}[!htbp] \centering 
	\caption{Coverage - CATE} 
	\label{tab:coverage.cate} 
	\small
	\begin{tabular}{@{\extracolsep{5pt}} cccccc} 
		\hline
		$\tau$ & $\delta_{\mu}$ & $\delta_{\tau}$ & BART-RDD & S-BART & T-BART \\ 
		\hline \\[-1.8ex] 
		$0.2$ & $0.5$ & $0.1$ & $0.993$ & $0.969$ & $0.951$ \\ 
		$0.2$ & $0.5$ & $0.3$ & $0.986$ & $0.904$ & $0.936$ \\ 
		$0.2$ & $1.25$ & $0.1$ & $0.985$ & $0.949$ & $0.828$ \\ 
		$0.2$ & $1.25$ & $0.3$ & $0.974$ & $0.897$ & $0.816$ \\ 
		$0.5$ & $0.5$ & $0.1$ & $0.986$ & $0.919$ & $0.955$ \\ 
		$0.5$ & $0.5$ & $0.3$ & $0.985$ & $0.933$ & $0.941$ \\ 
		$0.5$ & $1.25$ & $0.1$ & $0.980$ & $0.909$ & $0.820$ \\ 
		$0.5$ & $1.25$ & $0.3$ & $0.982$ & $0.922$ & $0.835$ \\ 
		\hline \\[-1.8ex] 
	\end{tabular} 
\end{table} 

\begin{figure}
	\centering
	\begin{subfigure}[b]{0.49\textwidth}
		\centering
		\includegraphics[width=\textwidth]{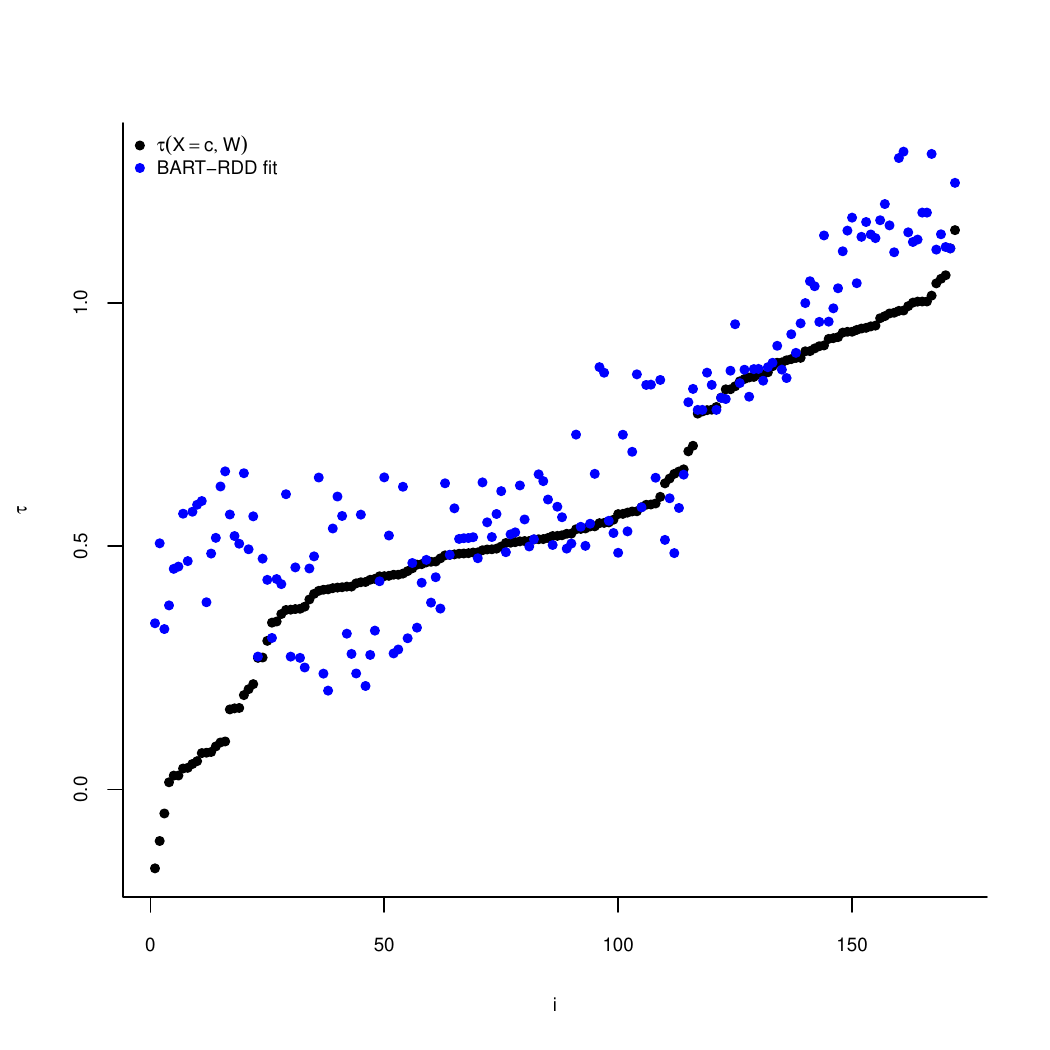}
		\caption{BART-RDD}
		\label{fig:cate.bart.rdd.1}
	\end{subfigure}
	\hfill
	\begin{subfigure}[b]{0.49\textwidth}
		\centering
		\includegraphics[width=\textwidth]{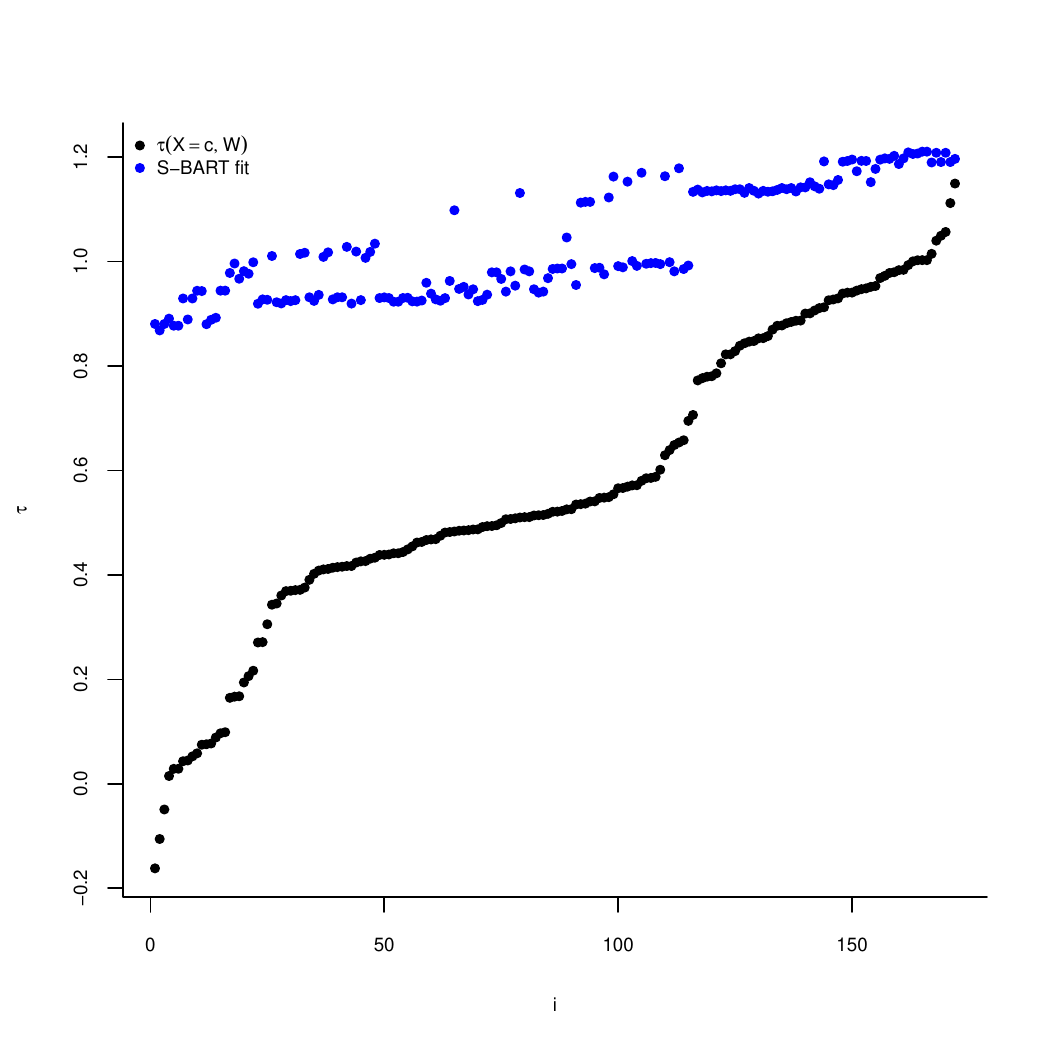}
		\caption{S-BART}
		\label{fig:cate.sbart.1}
	\end{subfigure}
	\hfill
	\begin{subfigure}[b]{0.49\textwidth}
		\centering
		\includegraphics[width=\textwidth]{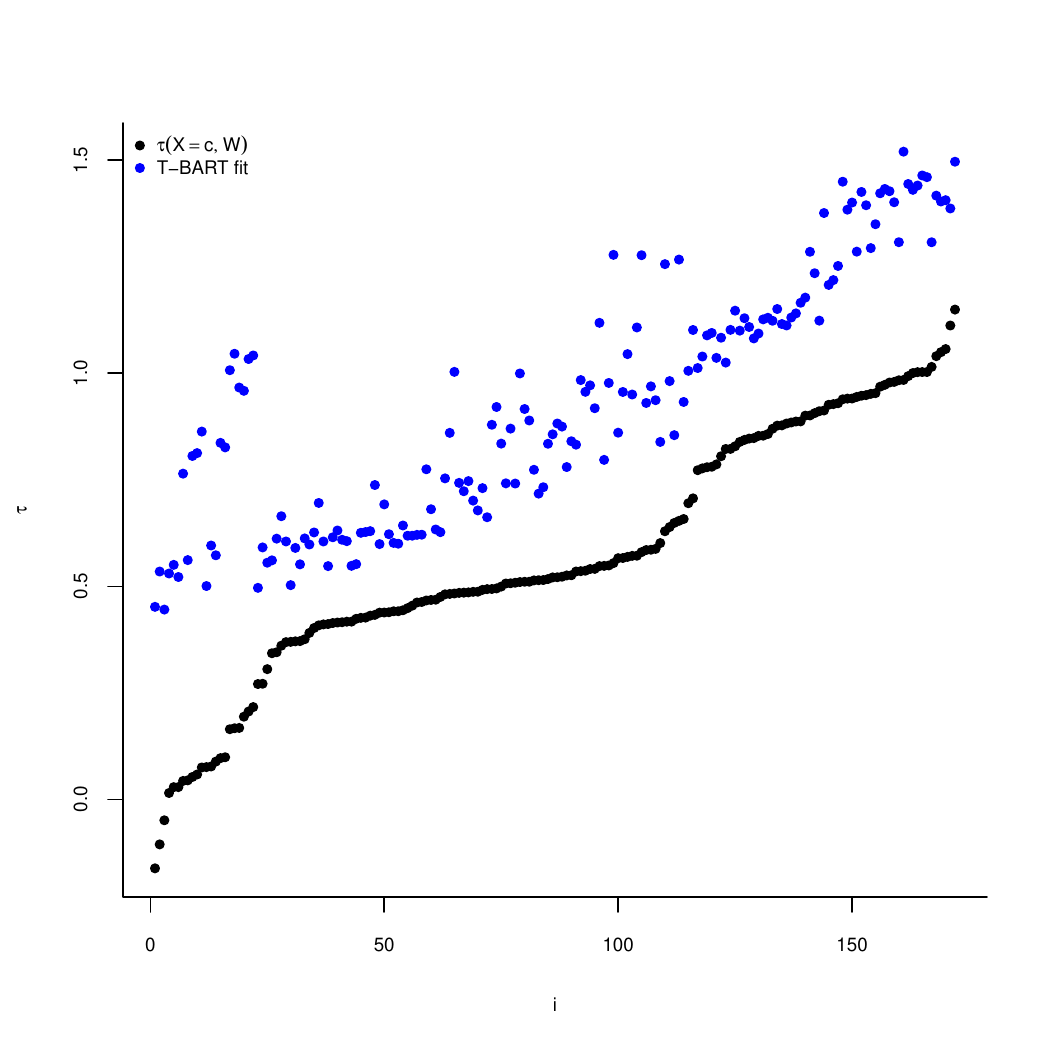}
		\caption{T-BART}
		\label{fig:cate.tbart.1}
	\end{subfigure}
	\caption{Fit for $\tau(X=c,W)$ for each method when $\delta_{\mu}=0.5$ versus the true function}
	\label{fig:cate.1}
\end{figure}

\begin{figure}
	\centering
	\begin{subfigure}[b]{0.49\textwidth}
		\centering
		\includegraphics[width=\textwidth]{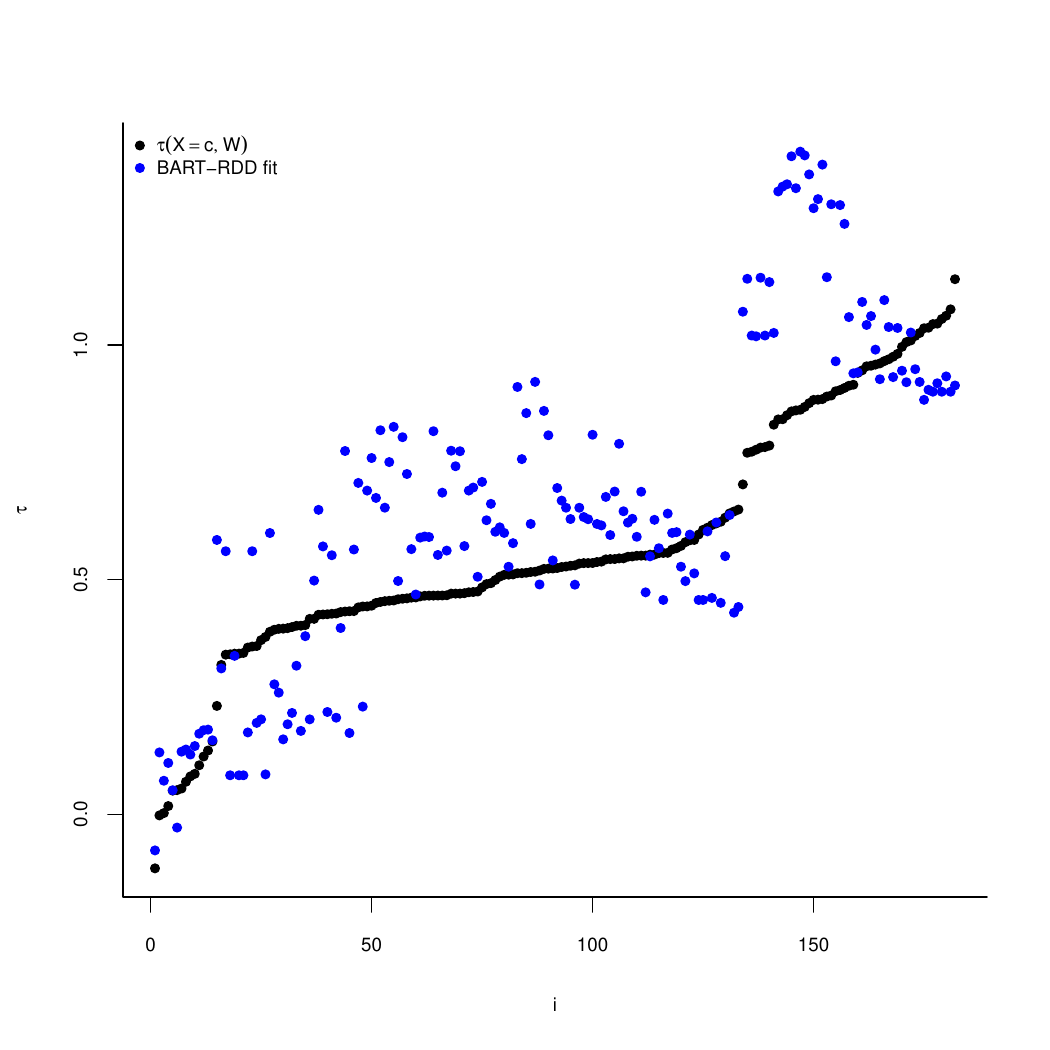}
		\caption{BART-RDD}
		\label{fig:cate.bart.rdd.2}
	\end{subfigure}
	\hfill
	\begin{subfigure}[b]{0.49\textwidth}
		\centering
		\includegraphics[width=\textwidth]{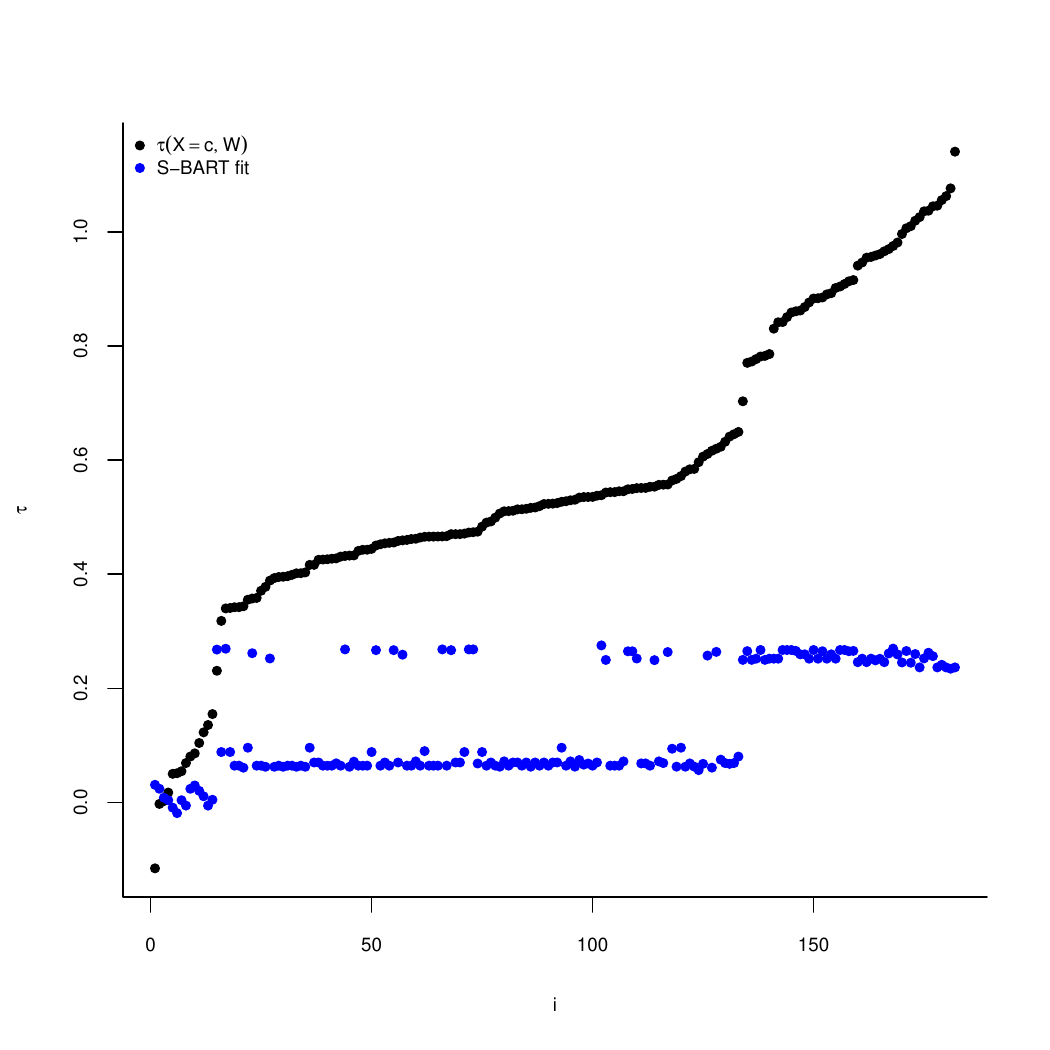}
		\caption{S-BART}
		\label{fig:cate.sbart.2}
	\end{subfigure}
	\hfill
	\begin{subfigure}[b]{0.49\textwidth}
		\centering
		\includegraphics[width=\textwidth]{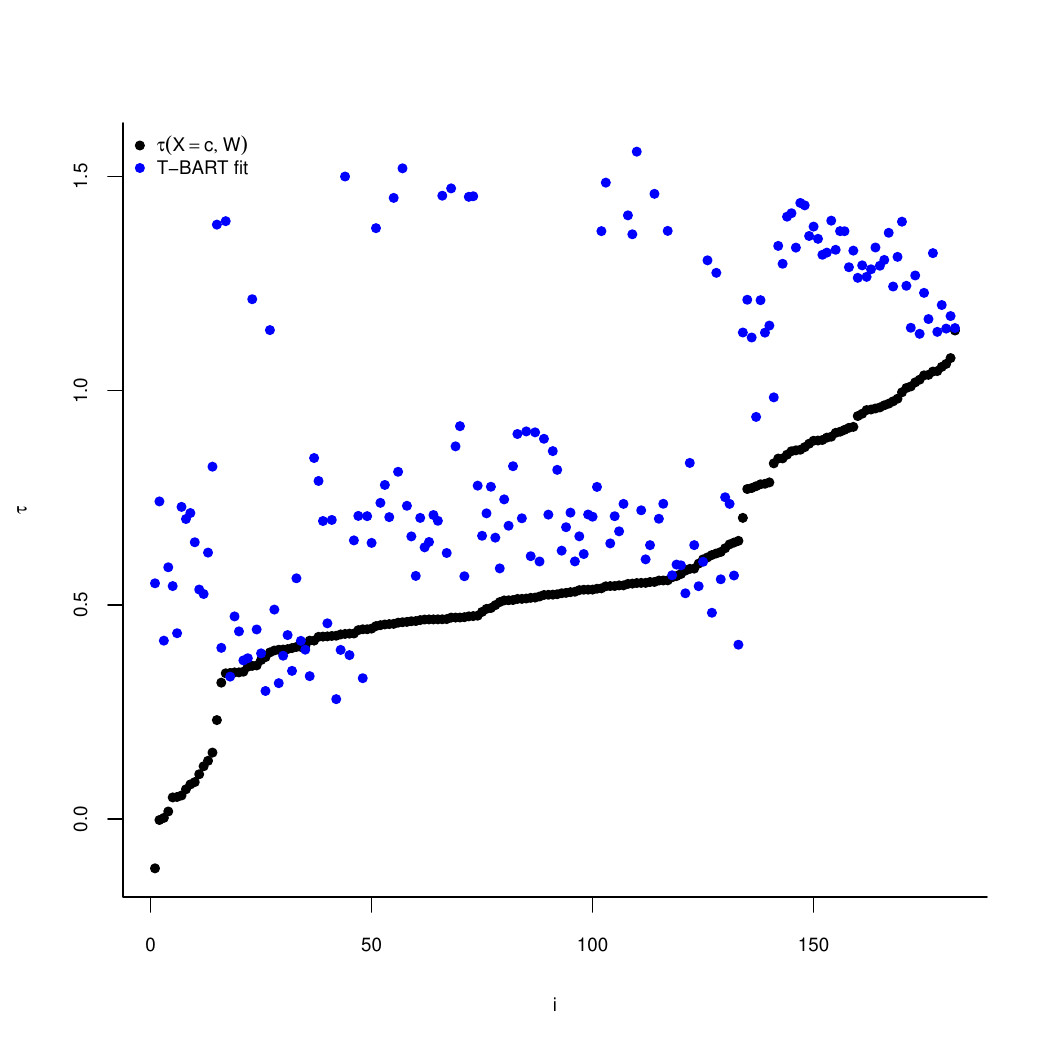}
		\caption{T-BART}
		\label{fig:cate.tbart.2}
	\end{subfigure}
	\caption{Fit for $\tau(X=c,W)$ for each method when $\delta_{\mu}=1$ versus the true function}
	\label{fig:cate.2}
\end{figure}

For a more detailed look into the CATE predictions of each model, figures \ref{fig:cate.1} and \ref{fig:cate.2} present the CATE fit for an illustrative sample of the DGP described earlier, with $\delta_{\mu}=0.5$ and $\delta_{\mu}=1.25$ respectively. We set $\delta_{\tau}=0.3$ and $\bar{\tau}=0.5$ for these examples. The values are presented for units inside the identification strip in ascending order. Two patterns stand out in these comparisons. First, although increasing $\delta_{\mu}$ evidently makes it harder to recover $\tau$ in general, the results from BART-RDD are a lot less sensitive to these changes. Second, S-BART seems to have a lot more difficulties in picking up variations in $W$, producing much more constant CATE estimates than the other methods. Overall, the figures suggest that the BART-RDD CATE predictions are less biased and more able to capture heterogeneity than the unmodified BART models.


\subsection{Summary of Simulation Results}
\label{sec:org395252e}

These simulation results bear out our motivating problem: unmodified BART models have a difficult time coping sensibly with the lack of overlap in RDDs. BART's lack of control over how the leaves containing points near the cutoff are formed can lead to nearly empty nodes in that region. This is especially true if the conditional expectations have a large spread in that region, either because of steepness in $X$ or because of strong heterogeneity in $W$, as this will probably lead to many splits. These issues arise even more prominently for the T-BART model, since this model only features observations from one side of the cutoff or the other by construction. The BART-RDD model avoids these issues by having direct BART priors for the prognostic and treatment effect components and restricting the growth process of these trees to ensure that nodes containing the cutoff point are well populated by points from both sides of the cutoff.

The effectiveness of BART-RDD in controlling the potential bias of unmodified BART models in the CATE estimation naturally carries over to the ATE estimates. BART-RDD produces ATE estimates that are generally better and never far worse than those produced by the polynomial estimators. Being a Bayesian model for $\E[Y \mid X,W]$, it is expected that BART will control variance at the cost of some bias in the ATE estimation compared to LLR. However, for the reasons discussed above, this bias might be stronger than the variance reduction in the unmodified BART models. BART-RDD, by controlling this bias, is able to inherit the good predictive capabilities of BART into this context, leading to competitive ATE estimates, even though the main focus is CATE estimation.

\section{The Effect of Academic Probation in Educational Outcomes}
\label{sec:application}
We now conclude with a detailed empirical application of BART-RDD to
illustrate its usage in a real data setting. The data
analyzed in this section comes from \citet{lindo2010ability}
and consists of information on college students enrolled in
a large Canadian university. Students who, by the end of
each term, present GPA lower than a certain threshold (which
differs between the three university campuses) are placed on
academic probation and must improve their GPA in the next
term and face threat of punishment if they fail to achieve
this goal, which can range from 1-year to permanent
suspension from the university.

Among the performance outcomes analyzed by
\citet{lindo2010ability}, we focus on GPA in the term after a
student is placed on probation (\(Y\)). Following the authors,
we define the running variable (\(X\)) as the negative
distance between a student's first-year GPA and the
probation threshold, meaning students below the limit have a
positive score and the cutoff is 0. Additional student
features in the data include gender (`male'), age when
student entered the university
(`age\_at\_entry'), a dummy for being born in
North America (`bpl\_north\_america'), attempted
credits in the first year (`totcredits\_year1'),
dummies for which campus each student belongs to
(`loc\_campus' 1, 2 and 3), and the student's position
in the distribution of high school grades of students
entering the university in the same year as a measure of
high school performance (`hsgrade\_pct').

\begin{figure}[!htpb]
	\centering
	\includegraphics[width=0.75\linewidth]{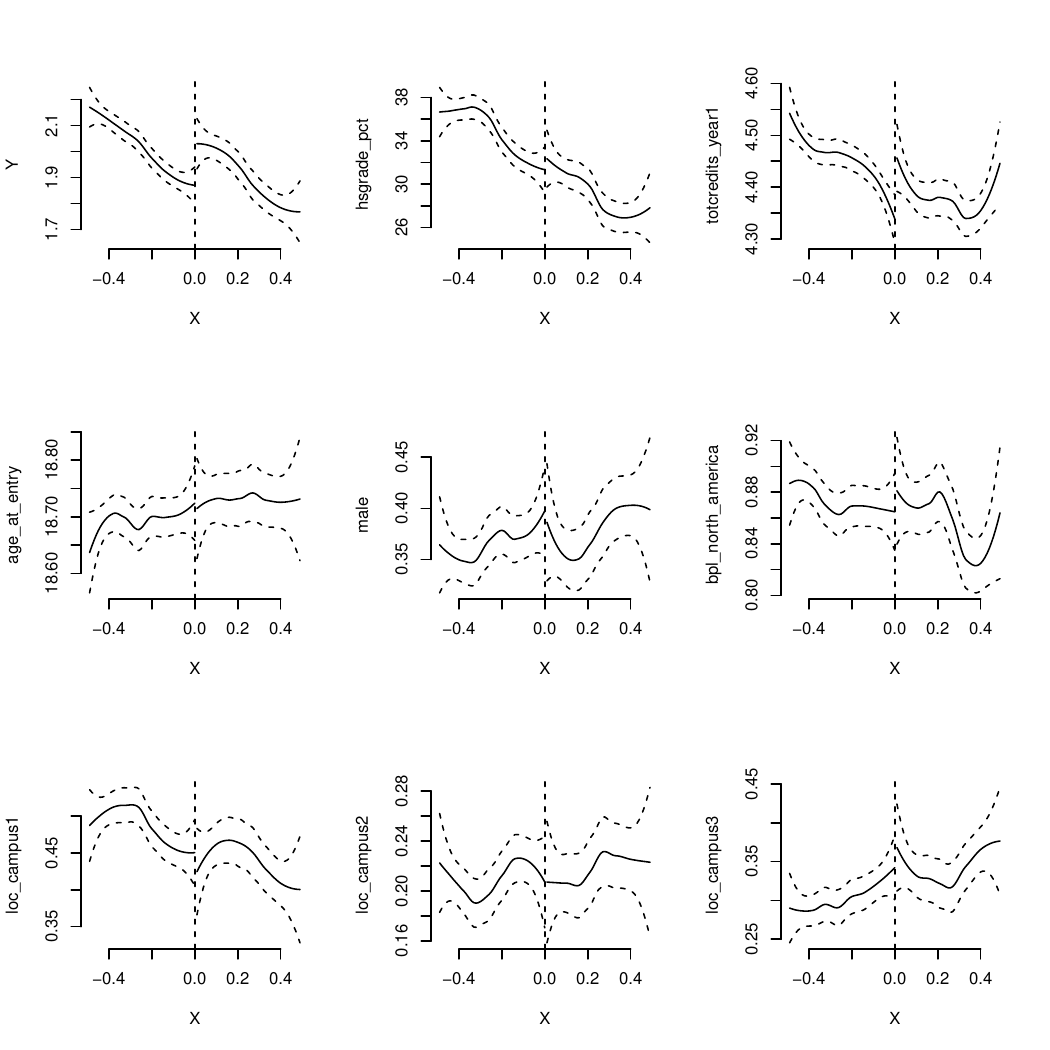}
	\caption{Outcome and covariate distribution conditional
		on $X$}
	\label{fig:application.summary}
\end{figure}

Figure \ref{fig:application.summary} presents a LOESS fit
with a 95\% confidence band for the distribution of \(Y\) and
each covariate conditional on \(X\) for \(X \in [-0.5,0.5]\). We
see a clear negative relationship between \(Y\) and \(X\),
meaning students who had a lower GPA in the first year are
more likely to have a lower GPA in the second year as
well. There is also a clear discontinuity at the probation
threshold. Among the covariates, only high-school grades and
total credits attempted in the first year have a clear
correlation with \(X\), both of them negative. This means that
students with low first-year GPA are also more likely to
have had bad high-school grades and to have attempted less
credits in the first year. The latter feature also presents
a discontinuity at the probation threshold, which means it must be included in the estimation to avoid bias.

Table \ref{tab:sum.stat} presents the mean, standard
deviation, minimum, maximum and correlation with \(Y\) for
each variable in the full sample and per campus. The running
variable, high-school grade percentile and credits attempted
seem to be the strongest predictors. In terms of campus
composition, the running variable and high-school
performance are the only ones with clearly varying
distributions across campus. The former feature presents a
lower mean for campus 1 compared to the other two, while the
latter presents a higher mean for campus 1. This means
students in campus 1 generally performed better in
high-school and obtain better GPA scores by the end of their
first year. As discussed by \citet{lindo2010ability}, the
campuses are indeed different in their student
composition. Campus 1 is the central campus and has a more
traditional university structure, lower acceptance rates and
more full-time students, while campuses 2 and 3 are
satellite campuses and resemble community colleges more,
with a higher acceptance rate and more part-time and
commuter students. These differences suggest not only that
the expected second-year GPA should differ across campuses,
but also that the probation policy could have differential
impacts between campuses.

\begin{table}[!htpb] \centering 
	\caption{Summary statistics} 
	\label{tab:sum.stat} 
	\small 
	\begin{tabular}{@{\extracolsep{5pt}} lcccccc} 
		\hline
		& Sample & Mean & SD & Min & Max & Cor \\ 
		\hline
		\multirow{4}{*}{Y} & Full & 2.571 & 0.91 & 0 & 4.3 & 1 \\ 
		& Campus 1 & 2.676 & 0.897 & 0 & 4.3 & 1 \\ 
		& Campus 2 & 2.486 & 0.886 & 0 & 4.3 & 1 \\ 
		& Campus 3 & 2.369 & 0.921 & 0 & 4.3 & 1 \\
		\hline
		\multirow{4}{*}{X} & Full & -0.961 & 0.864 & -2.8 & 1.6 & -0.656 \\ 
		& Campus 1 & -1.113 & 0.83 & -2.8 & 1.5 & -0.652 \\ 
		& Campus 2 & -0.79 & 0.84 & -2.8 & 1.5 & -0.64 \\ 
		& Campus 3 & -0.706 & 0.881 & -2.7 & 1.6 & -0.642 \\
		\hline
		\multirow{4}{*}{hsgrade\_pct} & Full & 51.003 & 28.712 & 1 & 100 & 0.47 \\ 
		& Campus 1 & 60.282 & 26.021 & 1 & 100 & 0.456 \\ 
		& Campus 2 & 37.165 & 27.057 & 1 & 100 & 0.488 \\ 
		& Campus 3 & 37.878 & 27.074 & 1 & 100 & 0.437 \\
		\hline
		\multirow{4}{*}{totcredits\_year1} & Full & 4.584 & 0.505 & 3 & 6.5 & 0.222 \\ 
		& Campus 1 & 4.694 & 0.435 & 4 & 6.5 & 0.191 \\ 
		& Campus 2 & 4.465 & 0.47 & 4 & 6 & 0.129 \\ 
		& Campus 3 & 4.395 & 0.609 & 3 & 6 & 0.237 \\
		\hline
		\multirow{4}{*}{age\_at\_entry} & Full & 18.656 & 0.735 & 17 & 21 & -0.09 \\ 
		& Campus 1 & 18.631 & 0.72 & 17 & 21 & -0.089 \\ 
		& Campus 2 & 18.658 & 0.717 & 17 & 21 & -0.061 \\ 
		& Campus 3 & 18.716 & 0.781 & 17 & 21 & -0.09 \\
		\hline
		\multirow{4}{*}{male} & Full & 0.38 & 0.485 & 0 & 1 & -0.039 \\ 
		& Campus 1 & 0.378 & 0.485 & 0 & 1 & -0.039 \\ 
		& Campus 2 & 0.363 & 0.481 & 0 & 1 & -0.055 \\ 
		& Campus 3 & 0.398 & 0.489 & 0 & 1 & -0.024 \\
		\hline
		\multirow{4}{*}{bpl\_north\_america} & Full & 0.87 & 0.337 & 0 & 1 & 0.02 \\ 
		& Campus 1 & 0.874 & 0.332 & 0 & 1 & 0.012 \\ 
		& Campus 2 & 0.897 & 0.305 & 0 & 1 & 0.02 \\ 
		& Campus 3 & 0.84 & 0.367 & 0 & 1 & 0.022 \\ 
		\hline
		\multicolumn{7}{l}{\tiny Sample size} \\ 
		\multicolumn{7}{l}{\tiny Total: 40582; Campus 1: 23999; Campus 2: 7029; Campus 3: 9554}
	\end{tabular} 
\end{table}

In order to determine the appropriate prior parameters for this sample,
we perform the prior elicitation procedure described in section \ref{sec:prior.elicitation}: fix \(X,W\), generate 10 samples of the DGP described in that section, take a grid of candidate values for $(N_{Omin},N_{Opct},h)$ and calculate the RMSE over the 10 synthetic samples for each combination in the grid\footnote{Because of the sample size of over 40,000 points, we are able to explore the prior reasonably with as few as 10 synthetic samples; for data with smaller sample sizes, more synthetic samples might be necessary to clearly distinguish between the candidates.}. Table \ref{tab:rmse.data} presents the results of this procedure. We set the parameters at the values which yield the lowest RMSE: $(N_{Omin},N_{Opct},h) = (5,0.6,0.1)$.

\begin{table}[!htbp] \centering 
	\caption{Results from prior elicitation} 
	\label{tab:rmse.data} 
	\small 
	\begin{tabular}{@{\extracolsep{5pt}} cccc} 
		\hline 
		$N_{Omin}$ & $N_{Opct}$ & $h$ & RMSE \\ 
		\hline
		$5$ & $0.6$ & $0.1$ & $0.094$ \\ 
		$5$ & $0.8$ & $0.15$ & $0.096$ \\ 
		$10$ & $0.8$ & $0.15$ & $0.107$ \\ 
		$10$ & $0.6$ & $0.1$ & $0.126$ \\ 
		$5$ & $0.7$ & $0.15$ & $0.247$ \\ 
		$10$ & $0.7$ & $0.15$ & $0.247$ \\ 
		$5$ & $0.6$ & $0.15$ & $0.252$ \\ 
		$10$ & $0.6$ & $0.15$ & $0.254$ \\ 
		$5$ & $0.7$ & $0.1$ & $0.331$ \\ 
		$5$ & $0.6$ & $0.05$ & $0.343$ \\ 
		$10$ & $0.8$ & $0.05$ & $0.347$ \\ 
		$5$ & $0.8$ & $0.05$ & $0.349$ \\ 
		$5$ & $0.7$ & $0.05$ & $0.353$ \\ 
		$10$ & $0.7$ & $0.05$ & $0.357$ \\ 
		$10$ & $0.7$ & $0.1$ & $0.358$ \\ 
		$10$ & $0.6$ & $0.05$ & $0.368$ \\ 
		$5$ & $0.8$ & $0.1$ & $0.391$ \\ 
		$10$ & $0.8$ & $0.1$ & $0.398$ \\ 
		\hline
	\end{tabular} 
\end{table} 

The model is fit for the whole sample but the treatment
effect function at the cutoff is predicted only for the
points inside the identification strip. Table
\ref{tab:sum.stat.h} presents summary statistics for this
prediction sample. Second-year GPA is consistently greater
for treated units overall and per campus. Besides that,
gender is the only feature that exhibits some difference
between treatment groups, with \(40.4\%\) untreated and \(26.4\%\)
treated men in campus 2, while the gender distribution for
campus 1 and 3 is similar across treatment groups. The only
feature that differs significantly across campuses is the
high-school grade percentile: campus 1 is composed of
students which had better high-school performance than those
of campus 2 or 3, which are similar in that
regard. Generally, the prediction sample presents a similar
feature distribution across treatment groups and campuses,
with the exception of high-school performance for campus 1
and gender for campus 2.

\begin{table}[!htpb] \centering 
	\caption{Summary statistics - identification strip} 
	\label{tab:sum.stat.h} 
	\small 
	\begin{tabular}{@{\extracolsep{5pt}} l|l|cccc}
		\hline
		& & \multicolumn{2}{c}{Control} & \multicolumn{2}{c}{Treatment} \\ 
		\hline
		& Sample & Mean & SD & Mean & SD \\ 
		\multirow{4}{*}{Y} & Full & 1.896 & 0.818 & 2.023 & 0.787 \\ 
		& Campus 1 & 1.931 & 0.847 & 2.007 & 0.808 \\ 
		& Campus 2 & 1.941 & 0.846 & 2.128 & 0.685 \\ 
		& Campus 3 & 1.823 & 0.756 & 1.976 & 0.819 \\ \hline
		\multirow{4}{*}{X} & Full & -0.042 & 0.031 & 0.051 & 0.026 \\ 
		& Campus 1 & -0.042 & 0.031 & 0.051 & 0.026 \\ 
		& Campus 2 & -0.041 & 0.031 & 0.053 & 0.025 \\ 
		& Campus 3 & -0.041 & 0.03 & 0.048 & 0.027 \\ \hline
		\multirow{4}{*}{hsgrade\_pct} & Full & 31.941 & 22.796 & 31.234 & 22.781 \\ 
		& Campus 1 & 42.354 & 22.233 & 43.041 & 22.402 \\ 
		& Campus 2 & 23.399 & 19.246 & 20.95 & 17.759 \\ 
		& Campus 3 & 23.494 & 19.698 & 22.475 & 18.668 \\ \hline
		\multirow{4}{*}{totcredits\_year1} & Full & 4.375 & 0.539 & 4.418 & 0.547 \\ 
		& Campus 1 & 4.494 & 0.459 & 4.588 & 0.446 \\ 
		& Campus 2 & 4.367 & 0.456 & 4.434 & 0.457 \\ 
		& Campus 3 & 4.225 & 0.638 & 4.184 & 0.633 \\ \hline
		\multirow{4}{*}{age\_at\_entry} & Full & 18.715 & 0.753 & 18.727 & 0.745 \\ 
		& Campus 1 & 18.708 & 0.731 & 18.701 & 0.712 \\ 
		& Campus 2 & 18.679 & 0.71 & 18.711 & 0.678 \\ 
		& Campus 3 & 18.746 & 0.806 & 18.773 & 0.826 \\ \hline
		\multirow{4}{*}{male} & Full & 0.387 & 0.487 & 0.362 & 0.481 \\ 
		& Campus 1 & 0.38 & 0.486 & 0.396 & 0.49 \\ 
		& Campus 2 & 0.404 & 0.492 & 0.264 & 0.442 \\ 
		& Campus 3 & 0.387 & 0.488 & 0.38 & 0.486 \\ \hline
		\multirow{4}{*}{bpl\_north\_america} & Full & 0.861 & 0.346 & 0.88 & 0.325 \\ 
		& Campus 1 & 0.865 & 0.342 & 0.865 & 0.342 \\ 
		& Campus 2 & 0.908 & 0.289 & 0.925 & 0.265 \\ 
		& Campus 3 & 0.828 & 0.378 & 0.872 & 0.335 \\ 
		\hline
		\multicolumn{6}{l}{\tiny Sample size (control/treatment):} \\ 
		\multicolumn{6}{l}{\tiny Total: 1038/719; Campus 1: 466/318; Campus 2: 218/159; Campus 3: 354/242} \\ 
	\end{tabular} 
\end{table} 

We generate 100 draws from the individual-level posterior
distribution which, averaging over observations, lead to 100 draws from the ATE posterior distribution. Table \ref{tab:ate.results} presents a summary of the ATE posterior. The distribution is centered at \(0.14\) with all the posterior mass above zero, indicating strong evidence for positive effects of the probation policy. The 95\% credible interval suggests the average effect can be as low as \(0.08\) and as high as \(0.217\)\footnote{For comparison, appendix section \ref{sec:application.others} presents the ATE results for the other estimators analyzed in the simulation exercise.}.

\begin{table}[!htbp] \centering 
	\caption{BART-RDD posterior summary for the ATE} 
	\label{tab:ate.results} 
	\small 
	\begin{tabular}{@{\extracolsep{5pt}} ccccccc} 
		\hline 
		Mean & SD & 2.5\% & 97.5\% & Median & Min & Max \\ 
		\hline  
		$0.140$ & $0.036$ & $0.080$ & $0.217$ & $0.140$ & $0.068$ & $0.253$ \\ 
		\hline  
	\end{tabular} 
\end{table} 

\begin{figure}[!htpb]
	\centering
	\includegraphics[width=0.75\linewidth]{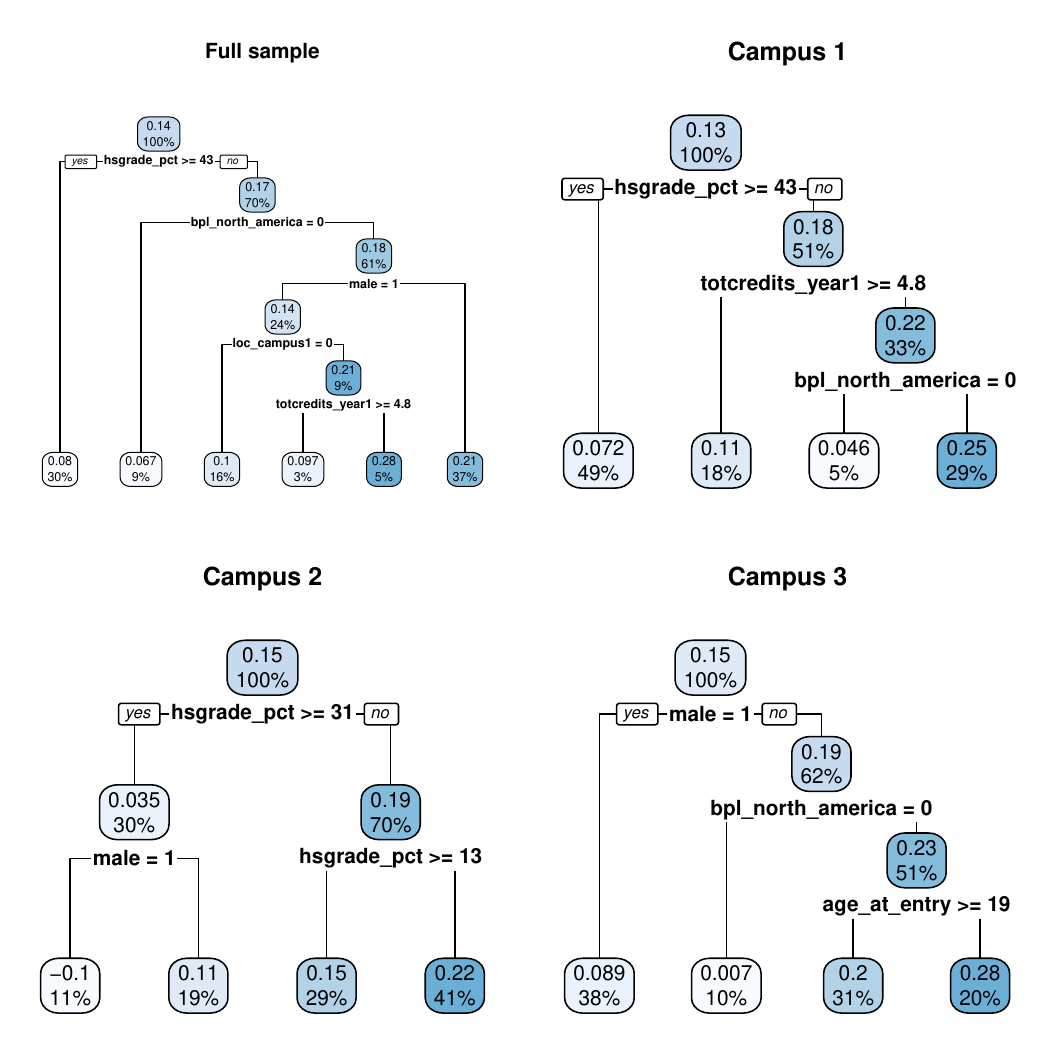}
	\caption{Regression tree fit to posterior point estimates
		of individual treatment effects: top number in each box is the average subgroup treatment effect, lower number shows the percentage of the total sample in that subgroup; the full sample summary flags high-school performance, birth place, gender, campus location and credits in first year as important moderators; the separate campus fits indicate heterogeneity between the campuses: for campus 1, high-school performance, credits attempted and birth place are flagged as important moderators, while for campus 2, high-school performance and gender are important and, for campus 3, gender, birth place and age at entry are the important moderators}
	\label{fig:cate.results}
\end{figure}

We now discuss heterogeneity in the BART-RDD posterior distribution. Figure \ref{fig:cate.results} presents a regression tree fit to posterior point estimates of the individual effects as a summarization tool. The summary trees are fit for the full sample and per campus. High-school performance is flagged as an important moderator for the full sample. Looking into each campus separately reveals more heterogeneity. For students who performed poorly in high-school in campus 1, we see additional moderation by birth place and credits attempted in the first year. In campus 2, we can see that the effect for women is larger than for men among those who feature above the 31-st percentile of high-school grades. Finally, for campus 3, the most important moderators are gender, birth place and age at entry.

The results described so far are consistent with those
presented by \citet{lindo2010ability}, both in magnitude and,
to some degree, in potential sources of treatment effect
heterogeneity. In particular, the authors also find a
greater effect for students who performed below average in
high-school and for women. Our posterior predictions however
flag additional features as potential moderators, such as
age at entry, birth place and campus, which highlights how depending on
pre-specification of relevant subgroups might lead
researchers to miss other interesting features of the
data. \citet{lindo2010ability} note that interpreting these
results as true effects requires caution since there is
evidence that the probation policy leads to differential
dropout rates, which changes the composition of students
before and after the evaluation of first-year GPA. However,
further discussion on this topic is out of the scope of this
project.

\begin{figure}[!htpb]
	\centering
	\includegraphics[width=0.75\linewidth]{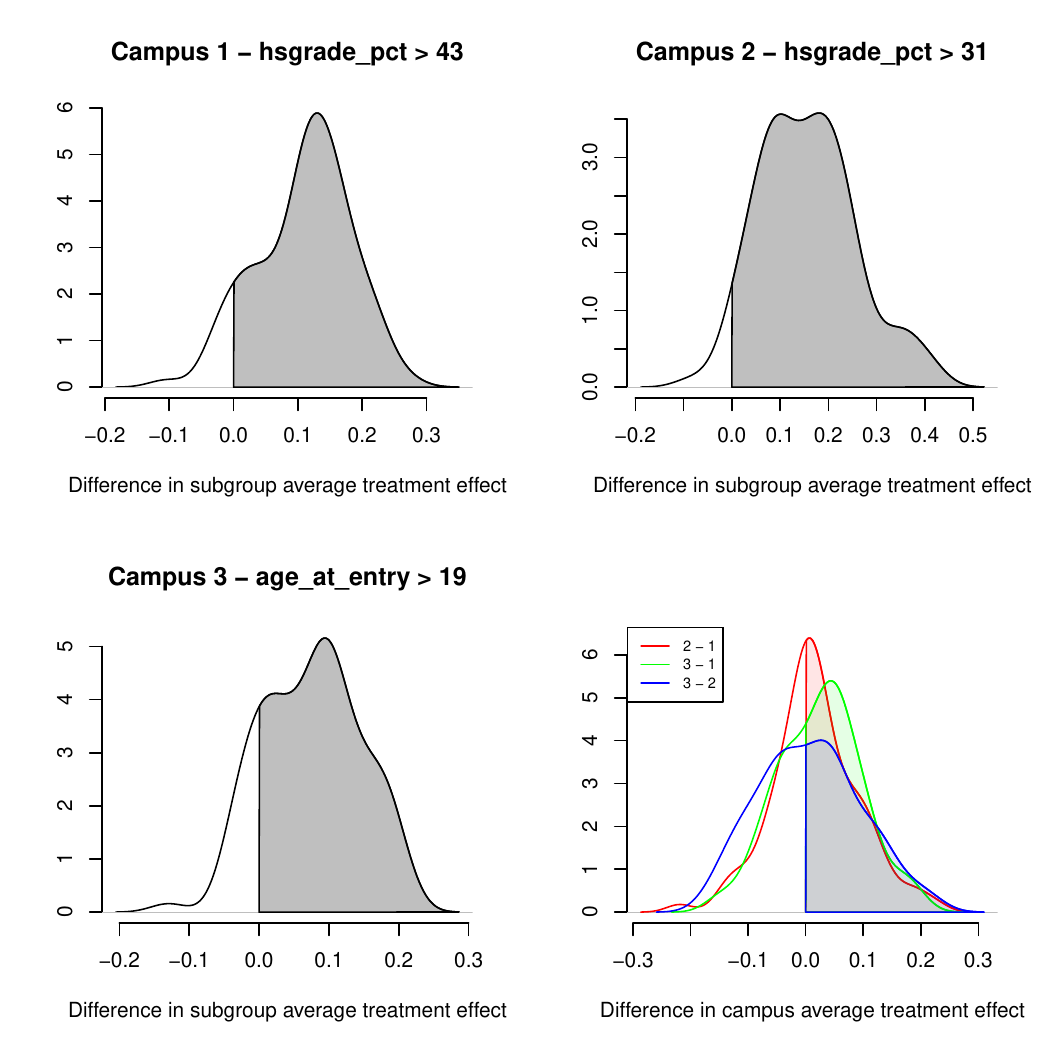}
	\caption{Differences in subgroup treatment effects: the first panel shows the posterior difference between students below and above the 43-rd percentile of high-school grades respectively in campus 1, which has a 92\% posterior mass above 0; the second panel performs the same analysis for the 31-st percentile of high-school grades for students in campus 2, which has a 95\% posterior mass above 0; the third panel presents the posterior difference between students that got into college younger versus older than 19 in campus 3, which has a posterior mass of 84\% above 0; the last panel presents the posterior differences in the ATE between each campus: there is a 66\% posterior probability of a larger effect for campus 3 compared to campus 1, a 59\% probability for a larger effect on campus 2 compared to campus 1 and a 54\% probability of a larger effect on campus 3 compared to campus 2}
	\label{fig:cate.differences}
\end{figure}

We conclude this section with an illustration of how to
perform posterior inference about heterogeneity in the
effects with the results of our model. Based on the
moderators flagged by the summarization trees, we
investigate the posterior difference in treatment effects
across some subgroups. The first panel in figure \ref{fig:cate.differences}
presents the posterior difference between students in the
bottom 43\% versus those in the upper 57\% of the high-school
grade distribution for campus 1. There is a 92\% posterior
probability that the treatment effect is larger for the
former group. The second panel presents a similar analysis
for campus 2, where the threshold was the 31-st percentile
of the high-school grade distribution. There is also strong
evidence for a larger effect for students lower in that
distribution, with a posterior probability of a larger
effect of 95\%. The third panel presents the posterior
difference for students who entered college younger than 19
versus those who entered older than that in campus 3. There
is also strong evidence of a larger effect for the former
group, with posterior probability of 84\%. Finally, the last
panel presents the posterior difference in average effects
between each campus. The biggest difference is between
campus 3 and campus 1, in which case there is a 66\%
probability of a larger effect for the former. There is a
59\% posterior probability that the effect is larger for
campus 2 than campus 1 and a 54\% posterior probability that
the effect is larger for campus 3 than campus 2.

\newpage
\bibliographystyle{apalike}
\bibliography{rdd_bart_arxiv.bib}
\appendix

\section{Prior elicitation experiments}\label{calibration}

Algorithm \ref{alg:prior} describes the procedure in pseudocode form. It is worth emphasizing that the particular functional form choices and parameter values for the synthetic data can be changed to better fit certain applications, although we recommend following the same general structure.

\begin{algorithm}
	\caption{Prior elicitation procedure}\label{alg:prior}
	\KwIn{Set of candidate prior parameter values $\Theta$, where $\theta \in \Theta$ is a 3-tuple $(h,\alpha,N_{Omin})$}
	\KwOut{Chooses $\theta^* \in \Theta$ to fit BART-RDD for the full sample $(Y,X,W)$}
	\KwData{Running variable and additional features, $(X,W)$}
	Generate $S$ samples of a synthetic outcome as follows:
	\begin{equation*}
		\begin{split}
			\mu(X,W) &= \frac{1}{P} \sum_{p=1}^P W_p + \frac{1}{1+\exp(-5X)}\\
			\tau(X) &= \bar{\tau} - \frac{\log(1+X)}{50}\\
			\bar{\tau} &= 0.4\\
			Y_s &= \mu(X,W) + \tau(X) Z + \varepsilon_s\\
			\varepsilon_s &\sim \N(0,1)
		\end{split}
	\end{equation*}
	
	\For{$\theta \in \Theta$}{\For{$s \in S$}{Fit BART-RDD for sample $(Y_s,X,W)$ with prior parameters $\theta$}\ Calculate $RMSE_{\theta}$ as the root-mean-square error of the ATE point estimates produced by BART-RDD for each of the S samples}\
	Choose $\theta^*$ which leads to the lowest $RMSE_{\theta}$
\end{algorithm}

\begin{figure}[!htpb]
	\centering
	\includegraphics[width=0.75\linewidth]{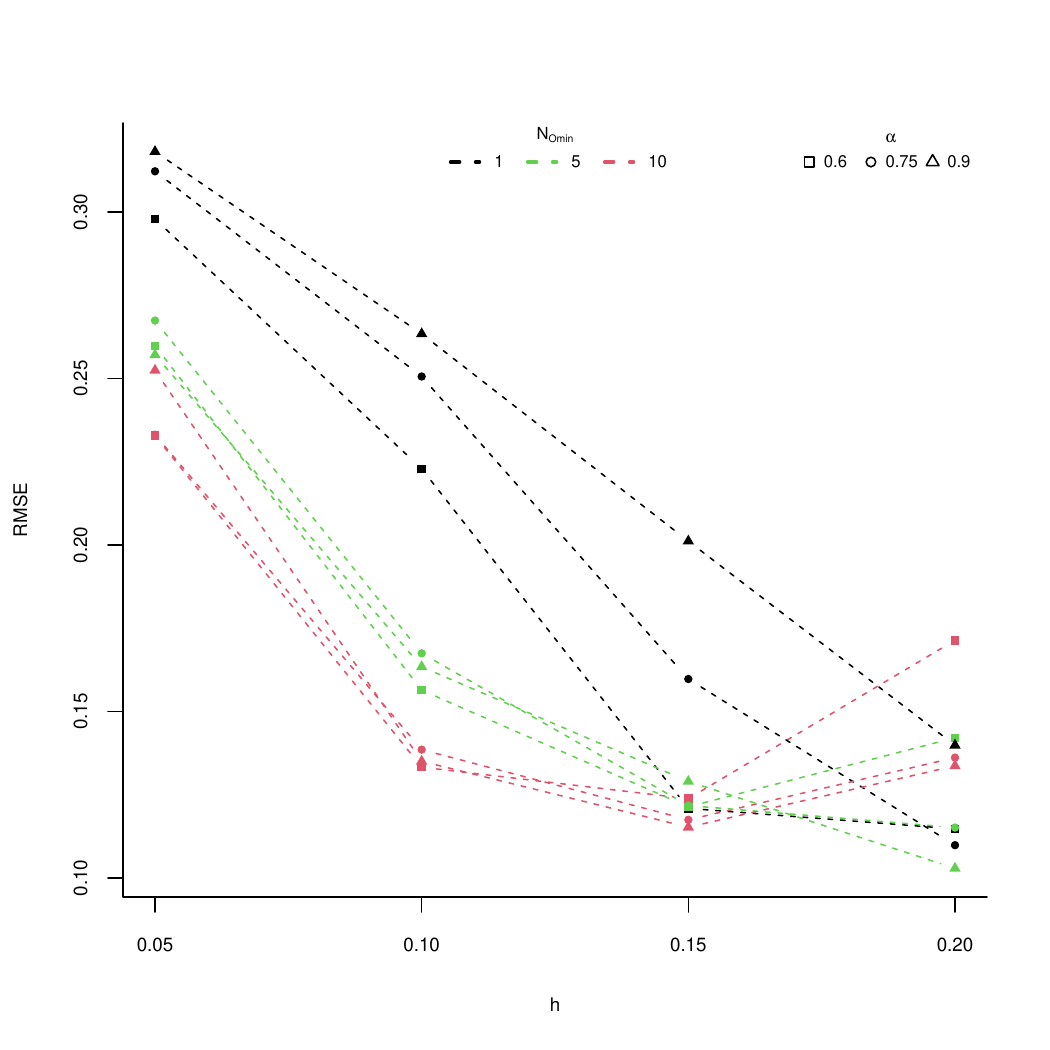}
	\caption{RMSE for each candidate $(h,\alpha,N_{Omin})$}
	\label{fig:prior.predictive}
\end{figure}

To illustrate this procedure, suppose we observe one sample
of \((X,W)\) of size 1000 from the DGP analyzed in the simulations\footnote{Note that, although there are several variations of the DGP considered in the simulations, the distribution of $(X,W)$ are always the same}. For that sample, we generate $S=20$ samples of
\(Y_s\) from a synthetic DGP constructed as described in algorithm \ref{alg:prior}. For the set of parameter candidates we consider $h \in \{ 0.05,0.1,0.15,0.2\}$, $N_{Omin} \in \{1,5,10\}$ and $\alpha \in \{0.6,0.75,0.9\}$. Figure \ref{fig:prior.predictive} presents the results of our exploration.

Although the RMSE patterns observed are specific to this sample, close inspection of figure \ref{fig:prior.predictive} allows us to observe some trends that can reasonably be expected to hold in many cases. Consider first the setup with small $h$. In this scenario, the nodes that include the identification strip are likely to be small unless the sample is well populated with points very close to the cutoff. One consequence is that, if we allow for nodes with only one point from each side of the cutoff in the strip (\textit{i.e.} $N_{Omin}=1$), we might end up with nodes that are very small, in which case our estimates are very unlikely to move away from the prior, which will generally lead to poor ATE estimates unless this parameter is also very small. Increasing $N_{Omin}$ safeguards against this possibility, as this would ensure the nodes are not that small. This is made clear by the fact that greater values for $N_{Omin}$ are uniformly better up until $h=0.1$.

Focusing next on the larger values of $h$, we see that $h=0.2$ is the setup that leads to the most sensitivity of the prior to the other parameters. Graphically, we see that the lines for each value of $N_{Omin}$ are less `clumped' together than for the lower values of $h$, meaning the combinations of $(\alpha,N_{Omin})$ lead to more varied results now. Practically, greater $h$ values mean we are using points that are potentially too far away from the cutoff for the ATE predictions at that point, which could evidently bias the results. In that case, lower $\alpha$ means we allow even more points far from the cutoff to influence these predictions, since we obtain nodes that may contain many points far from the identification strip. This can be seen in the figure since, for $h=0.2$, greater values of $\alpha$ generally produce the best results.

Overall, figure \ref{fig:prior.predictive} suggests setting $h$ and $N_{Omin}$ appropriately is crucial, while $\alpha$ can be set to offset any bias that might occur for a given combination of the other parameters. For this example, setting $N_{Omin}=10$ and $h = \{0.1,0.15\}$ seems reasonable. Then, we can set $\alpha$ accordingly. For example, if we choose $h=0.1$, setting $\alpha=0.6$ is best, although the other values lead to very similar results. In comparison, if we set $h=0.15$, greater $\alpha$ is uniformly better, as this value of $h$ seems to be large enough for this sample that we should trim the outter regions of the identification strip more strictly. However, the results are again not sensitive to $\alpha$ when $h=0.15$. Finally, it is worth noting that, although the lowest RMSE is achieved with $(h,\alpha,N_{Omin})=(0.2,5,0.9)$, we advise against setting $h=0.2$ for this sample given the much greater sensitivity of the prior to the other parameters in this case. Since one can never really know how close this synthetic DGP is to the real one, the search for the lowest RMSE here should be moderated by considering the sensitivity of the prior as well. With these considerations in mind, we suggest setting $(h,\alpha,N_{Omin})=(0.1,10,0.6)$ for this sample, which is the setup we use for the simulations.

\section{Application results for other estimators}\label{sec:application.others}

This section presents the ATE estimates for the \cite{lindo2010ability} data produced by the estimators studied in our simulations. S-BART and T-BART present evidence of a
near zero effect and the polynomial estimators
suggest a similar effect magnitude to BART-RDD. It is worth
noting that BART-based models have a regularization
component, which could explain
why the predictions from this models are more conservative
than those of the polynomial estimators (although only
slightly so for BART-RDD).

\begin{table}[!htbp] \centering 
	\caption{ATE point estimate and 95\% confidence interval for different estimators} 
	\label{tab:ate.others} 
	\small 
	\begin{tabular}{@{\extracolsep{5pt}} ccccc}  
		\hline 
		BART-RDD & S-BART & T-BART & LLR & CGS \\ 
		\hline 
		0.140 & 0.074 & 0.062 & 0.205 & 0.176 \\ 
		(0.080,0.217) & (-0.013,0.129) & (0,0.117) & (0.127,0.282) & (0.019,0.323) \\ 
		\hline
	\end{tabular} 
\end{table}
\end{document}